\documentclass[namedreferences,hyperref,optionalrh,solaromanenum]{spr-sola}

\usepackage{graphicx}                    
\usepackage{color}                       
\usepackage{xcolor}                       
\usepackage[separate-uncertainty = true,multi-part-units=single,range-units=single]{siunitx}
\usepackage{amssymb}                    
\usepackage{lineno}                     

\usepackage[english]{babel}
\usepackage[autostyle, english = american]{csquotes}
\MakeOuterQuote{"}

\chardef\us=`\_

\begin{document}

\begin{frontmatter}

\title{Thermal non-equilibrium cycles in a non-eruptive pseudo-streamer}


\author[addressref={aff1},corref,email={clara.froment@cnrs-orleans.fr}]{\inits{C.}\fnm{Clara}~\snm{Froment}\orcid{0000-0001-5315-2890}}
\author[addressref={aff2,aff3}]{\inits{S.}\fnm{Sophie}~\snm{Masson}\orcid{0000-0002-6376-1144}}

%
\runningauthor{Froment et al.}
\runningtitle{TNE cycles in a pseudo-streamer}

\address[id=aff1]{LPC2E, OSUC, Univ Orleans, CNRS, CNES, F-45071 Orleans, France}
\address[id=aff2]{Sorbonne Université, École polytechnique, Institut Polytechnique de Paris, Université Paris Saclay, Observatoire de Paris, Université PSL, CNRS, Laboratoire de Physique des Plasmas (LPP), Paris, France}
\address[id=aff3]{Observatoire Radioastronomique de Nançay, Observatoire de Paris, CNRS, PSL, Université d’Orléans, Nançay, France}

\begin{abstract}

Thermal non-equilibrium (TNE) is a well-known thermodynamic mechanism, generally studied in coronal loops. These evaporation and condensation cycles are induced by a quasi-steady and stratified heating. Long-period EUV pulsations and coronal rain are two manifestations of TNE. The quasi-periodicity of the cycles is a strong characteristic of TNE as the system dynamically evolves around a thermal equilibrium position that is not reachable. 
There are recent reports of coronal rain at open-closed boundaries such as fan-spine topologies and pseudo-streamers. However, no definite conclusions were drawn on the physical mechanisms driving these coronal rain events. In this paper, we report the detection of long-period EUV pulsations and co-spatial recurring coronal rain showers in a pseudo-streamer observed with SDO/AIA. The magnetic topology and evolution of the pseudo-streamer are studied in detail using a combination of PFSS modeling and EUV dynamics. The 2.5-day event exhibits all the characteristic features of previous TNE events reported in coronal loops: periodic EUV pulses appearing sequentially in the different channels, following the ordering of the peaks in their temperature response, and coronal rain showers appearing toward the end of these cycles.
We further show that the TNE cycles in the pseudo-streamer occur not only in the closed field but also in the open field. Our observations support the findings of recent numerical studies showing that TNE can also occur in open field regions and at open-closed boundaries. In parallel, interchange reconnection occurs continuously and non-impulsively all along the open-closed boundary as seen in EUV. The interplay between TNE and interchange reconnection may play a role in the release of condensations below the open-closed boundary and should be studied in detail in future works. 
This observation opens further perspectives for the understanding of TNE in the solar atmosphere and its potential implications for the solar wind.

\end{abstract}

%

\end{frontmatter}
%
\section{Introduction}\label{s:intro} 
Thermal non-equilibrium (TNE) is a counter-intuitive thermodynamic state, in which a heating distribution that is stable but imbalanced in altitude \citep{serioClosedCoronalStructures1981}, can produce a highly-dynamical behavior of the plasma in the solar atmosphere. 
The system, which couples the chromosphere to the corona, evolves dynamically --exchanging mass and energy--around an unreachable equilibrium position \citep{kuinThermalStabilityHot1982, martensCoolCoronalLoops1982}. The plasma will undergo cycles of evaporation (in the chromosphere) and condensation (in the corona) due to a heating mainly concentrated at low altitudes, and under quasi-steady conditions i.e., with a heating timescale that is shorter than the radiative cooling timescale \citep{antiochosModelFormationSolar1991}. 
The formation of coronal condensations is driven by a local excess of mass, brought in altitude by the chromospheric evaporation, whose coronal temperatures cannot be maintained by the local energy input. The plasma thus cools down, radiative losses increase, which accentuates the cooling, which in turn increases the radiative losses. A local thermal instability (TI) is triggered \citep{parkerInstabilityThermalFields1953, fieldThermalInstability1965}.
We refer readers to recent papers and reviews that address the mechanisms of TNE and TI, as well as their relationships, differences, and similarities, in detail \citep[e.g.,][]{klimchukDistinctionThermalNonequilibrium2019, klimchukRoleAsymmetriesThermal2019, antolinThermalInstabilityNonequilibrium2019, antolinMultiScaleVariabilityCoronal2022, watersSaturationMechanismThermal2023, watersCatastrophicCoolingInstability2025, keppensHydrodynamicThermalContinuum2025}.

The TI and TNE manifestations are widespread in the solar atmosphere. The TI-TNE scenario is one of the leading theories to explain the formation and dynamics of prominences \citep[e.g.,][]{antiochosModelFormationSolar1991, dahlburgProminenceFormationLocalized1998, antiochosDynamicFormationProminence1999, karpenAreMagneticDips2001, xiaFormationSolarFilaments2011, lunaFormationEvolutionMultithreaded2012, jercicProminenceCoronalRain2024} and coronal rain \citep[e.g.,][]{mullerDynamicsSolarCoronal2003, mullerDynamicsSolarCoronal2004, antolinMultithermalMultistrandedNature2015, moschouSimulatingCoronalCondensation2015, xiaCoronalRainMagnetic2017, liCoronalRainRandomly2022}.
Cool ($\sim 10^4$ to $10^5$ K) and dense ($\sim 10^{10}$ to $10^{12}$ cm$^{-3}$) blob-like features form in the corona and fall along coronal loops strands \citep[e.g.,][]{kawaguchiObservedInteractionProminences1970, leroyEmissionsFroidesDans1972, ahnActiveRegionCoronal2014, antolinMultithermalMultistrandedNature2015, kohutovaAnalysisCoronalRain2016}. 
TNE is also the main theory (the only one so far, to our best current knowledge) that can explain the properties of 
long-period extreme ultraviolet (EUV) pulsations \citep[][]{auchereLongPeriodIntensityPulsations2014, fromentEvidenceEvaporationincompleteCondensation2015, auchereThermalNonequilibriumRevealed2016, fromentLongperiodIntensityPulsations2017, pelouzeSpectroscopicDetectionCoronal2020, pelouzeRoleAsymmetriesCoronal2022}. These are detected in all EUV bands, have periods ranging from 2 to 16~hours and were estimated to occur (under several criteria) in at least half of the observed active regions \citep{auchereLongPeriodIntensityPulsations2014} and very remarkably in coronal loops (at least 30~\% of the detected cases).
In recent years, long-period EUV pulsations and cyclic coronal rain showers have been detected together \citep[][]{auchereCoronalMonsoonThermal2018, fromentMultiscaleObservationsThermal2020, sahinSpatialTemporalAnalysis2023}. The fluctuations around coronal temperatures produce EUV pulsations, while the rain is produced toward the end of the cooling phase of the TNE cycles. This behavior was recently reproduced self-consistently in 3D radiative magneto-hydrodynamics numerical simulations \citep[][]{luPeriodicCoronalRain2024}.

TNE has indeed received an increasing attention in the past decade, supported by new high-resolution and multi-thermal observations of the solar atmosphere, the discovery of long-period EUV pulsations and state-of-the-art numerical studies. Possible TNE manifestations in open coronal fields have sparked recent interest, despite being relatively unexplored thus far.
Some of these events fall under the category of hybrid prominence-coronal-rain events \citep{liuIRISObservationsNovel2016} whose magnetic structure involves open-closed magnetic field boundaries. These have been previously named "coronal spiders" or "cloud prominence" \citep[e.g.,][]{schadVectorMagneticField2016}, see \citet{antolinMultiScaleVariabilityCoronal2022} for an extended discussion. 
However, their connection to the TNE theory--which assumes quasi-constant and stratified heating that intrinsically produces cyclic cooling events--has yet to be demonstrated. Indeed, the condensations-- that is, coronal rain showers--are driven by TI and, in theory, could occur independently of TNE cycles.
Interchange reconnection, in which magnetic connectivity between closed and open fields is exchanged \citep[e.g.,][]{fiskBehaviorOpenMagnetic2001, edmondsonInterchangeReconnectionCoronal2010}, has been put forward as an additional or unique mechanism to initiate TI for these kinds of events.

This was first reported in \citet{liCoronalCondensationsCaused2018} and then in associated studies, exploring similar events \citep{liQuasiperiodicFastPropagating2018, liRepeatedCoronalCondensations2019, liRelationCoronalRain2020, chenCoronalCondensationSource2022a,qiaoThreeTypesSolar2024}. 
Using imaging, and for some of the studies spectroscopic observations, the authors described these events as being caused by reconnection between downward-moving open structures and low-lying loops. Magnetic dips would form in the higher-lying open structure in which coronal plasma accumulates. The local enhancement of density enables the TI-driven formation of condensations in the dips forming a small prominence-like feature. 
Some of the events were found to be recurring at the same location for several days \citep{liRepeatedCoronalCondensations2019}.

Contemporaneously, \citet{masonObservationsSolarCoronal2019} reported Raining Null Point Topologies (RNPTs) with the Atmospheric Imaging Assembly \citep[AIA,][]{boernerInitialCalibrationAtmospheric2012, lemenAtmosphericImagingAssembly2012} on board the Solar Dynamics Observatory \citep[SDO;][]{pesnellSolarDynamicsObservatory2012}, as also later reported in \citet{kumarPseudostreamerJetsCoronal2021}. 
These RNPTs are related to large null-point topologies with an open spine \citep{edmondsonInterchangeReconnectionCoronal2010}, and pseudo-streamers
\citep{titovMagneticTopologyCoronal2011}.
RNPTs are described as quite frequent and to happen in proximity of coronal holes. The rain showers seem to form preferentially in the open field (outer spine and near the null) but also in the closed field inside the dome, and to be recurrent. 
The authors discuss TNE and interchange reconnection as two viable mechanisms to explain their observations, individually or coupled. On the one hand, recurrent showers could be consistent with the TNE scenario. On the other hand, the authors argue that TNE is not expected to occur in open structures, as the radiative losses would be negligible compared to the  enthalpy flux outward. They propose that interchange reconnection is the mechanism that triggers the rain formation in the newly open field, filled with hot and dense plasma coming from the closed loops. The plasma in the open flux tube along the outer spine would be thus prone to TI.
We note many similarities between both types of events \citep{liCoronalCondensationsCaused2018,masonObservationsSolarCoronal2019} invoking both interchange reconnection and the TI-driven formation of condensations near nulls. By investigating further, we noticed that the recurring coronal rain events reported in \citet{liCoronalCondensationsCaused2018, liQuasiperiodicFastPropagating2018, liRepeatedCoronalCondensations2019} and that is  located on the west limb, at the border of the northern coronal hole on January 19, 2012, are actually happening in a pseudo-streamer which dynamics was studied in \citet{massonDynamicsTransitionCorona2014}. 

The motivation behind this present paper is to determine whether TNE cycles can be present at open-closed topologies. The discovery of long-period EUV pulsations unlocked a new way to detect and study TNE cycles. So far, this aspect is absent from all the previous works on coronal rain at open-closed topologies. Identifying the cyclic thermodynamical behavior, that is a key characteristic of TNE, would demonstrate that these hybrid prominence-coronal-rain events are part of the extended TNE events family.

In this paper, we report the detection of long-period EUV pulsations in a pseudo-streamer topology. The properties of these pulsations and their synchronicity with coronal rain showers are typical of TNE cycles. Indeed, these events show strong similarities with long-period EUV pulsation events studied extensively \citep[e.g.,][]{fromentEvidenceEvaporationincompleteCondensation2015, auchereCoronalMonsoonThermal2018, fromentMultiscaleObservationsThermal2020, sahinSpatialTemporalAnalysis2023} in the most well-know topology for TNE, that is, coronal loops. In Section~\ref{s:PS_context}, we introduce the datasets, describe the long term evolution of the EUV structure and study the underlying magnetic topology. We then dive into the analysis of long-period EUV pulsations and report on the signatures of TNE in Section~\ref{s:TNE_signatures}. In Section~\ref{s:PS_dynamic_and_TNE} we connect the TNE cycles observations with the pseudo-streamer dynamics and topology.
Finally, in Section~\ref{s:Conclusion}, we summarize our findings and their interpretations, and discuss their limitations and implications. 

\section{Observations of a non-eruptive pseudo-streamer}\label{s:PS_context} 

\subsection{Global evolution of the EUV emission in the \SI{171}{\angstrom} AIA channel}\label{ss:euv_PS_observation}

The observations analyzed in this paper were taken in June 2012 by SDO/AIA. We track the evolution of a structure that appears at the southeast limb from June 14, 2012 00:00 UT to June 16, 2012 12:00 UT. These 2.5 days cover most of the appearance and then disappearance of the structure in the AIA channels. 

We use the six coronal channels of AIA: \SI{94}{\angstrom}, \SI{131}{\angstrom}, \SI{171}{\angstrom}, \SI{193}{\angstrom}, \SI{211}{\angstrom} and \SI{335}{\angstrom}. We also use the \SI{304}{\angstrom} channel, whose temperature response has two components, peaking at coronal and transition region temperatures, respectively, and is thus well suited for the analysis of coronal rain.
Since we aim to focus on hours-long EUV pulsations and not on the shorter term dynamics, we limit the AIA cadence to \SI{10}{\minute} for most of the analysis. This dataset will be referred to as the "pulsation dataset". As explained later in Section~\ref{ss:pulsations_analysis}, this dataset will be sometimes spatially binned $4\times4$. Images that have an exposure time below \SI{0.5}{\second} are discarded. This represents only 6 and 15 images  (among 360 timesteps) for the \SI{171}{\angstrom} and \SI{193}{\angstrom} channels, respectively. For the study of coronal rain showers, we use observations at a cadence of \SI{1}{\minute}. This dataset will be referred to as the "rain dataset". For this dataset, we will always keep the original spatial resolution.
The data are calibrated and normalized to exposure time using the \texttt{aiapy} and \texttt{SunPy} \citep{barnesSunPyProjectOpen2020} routines. 
{Throughout the paper, we will apply additional processing to these datasets, as described in the relevant sections.

\begin{figure} 
\centerline{\includegraphics[width=\textwidth,clip=]{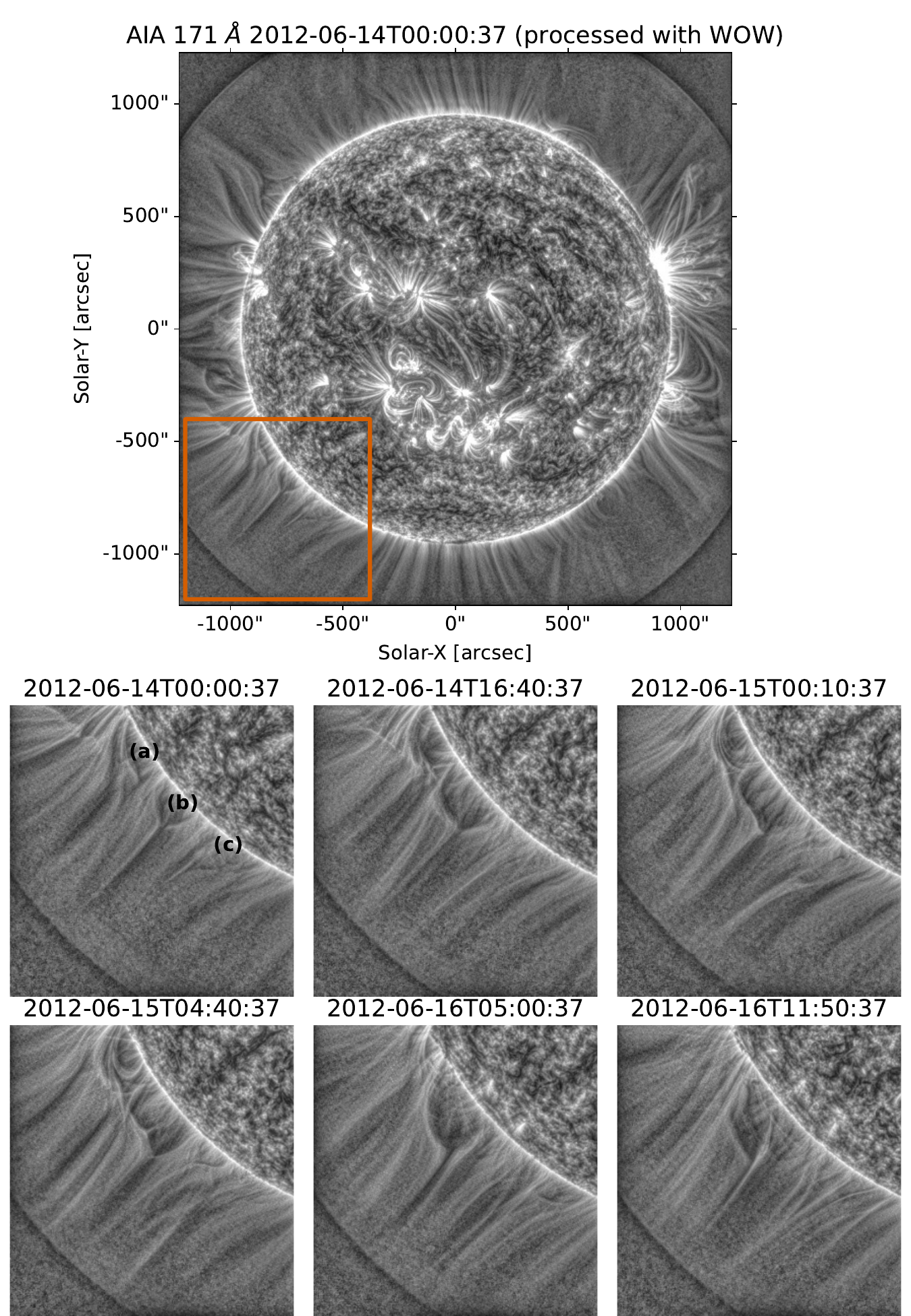}}
\caption{ROI observed by SDO/AIA in its \SI{171}{\angstrom} channel. All images are processed with the WOW algorithm \citep{auchereImageEnhancementWaveletoptimized2023}. Top: Full-disk image with the ROI delimited by an orange box $\SI{800}{\arcsecond}\times\SI{820}{\arcsecond}$, centered on (X,Y) = (\SI{-800}{\arcsecond},\SI{-790}{\arcsecond}). Bottom: Temporal evolution of the pseudo-streamers in the ROI. In the first snapshot we indicate the northern (a), middle (b), and southern (c) lobes which evolution is described in Section~{\ref{ss:euv_PS_observation}}.}
\label{fig:fig1_context}
\end{figure}

In Figure~\ref{fig:fig1_context}, we present the region of interest (ROI) we cut into the full-disk images. For this visualization, we use the wavelet-optimised whitening algorithm \citep[WOW;][]{auchereImageEnhancementWaveletoptimized2023}. We will use WOW throughout the paper to highlight the evolution of the structure tracked and to point to likely magnetic field configuration as deciphered from the \SI{171}{\angstrom} AIA channel emission, where it is best observed. We note that here we use spatially binned "pulsation dataset" but this has no impact on the visualization.
We identify closed magnetic structures forming three lobes, embedded in apparent open magnetic field regions, formed by true open field region or large closed loops with an apex located outside of the field-of-view (FOV) of AIA. Such EUV structures are usually associated with null point and pseudo-streamer topologies \citep[e.g.,][]{wangCoronalPseudostreamers2007, seatonSWAPFilterSimple2023, massonDynamicsTransitionCorona2014}. At the limb, those three lobes are aligned along the north-south} axis and are defined as the northern (a), middle (b), and southern (c) lobes. 
Looking at LASCO/C2 images with JHelioviewer \citep{mullerJHelioviewerTimedependent3D2017}, we find that the open EUV structure associated with the  middle lobe extends  outward a few solar radii.

We pick times at which there were milestones in the apparent modification of the magnetic field, as guessed from the EUV structures evolution. These form the temporal sequence shown in Figure~\ref{fig:fig1_context}.
At the beginning of the sequence, on June 14, 2012 00:00 UT, the three EUV lobes are well defined and separated from each others by some apparent open field. With time, the northern and middle lobes seem to merge (snapshots on June 14, 2012 16:40 UT, on June 15, 2012 00:10 UT, and on June 15, 2012 04:40 UT), and a final EUV lobe forms (snapshot on June 15, 2012 05:00 UT). In the snapshot on June 16, 2012 11:50 UT this lobe almost disappears at \SI{171}{\angstrom}. As the apparent merging occurs, the supposedly open-field region located between the northern and the middle lobes disappears. The temporal sequence studied here covers 60 hours of observation, meaning that the EUV structures observed at the limb travel a distance of about $600~\rm{Mm}$, equivalent to roughly a $\frac{1}{10}$ of a full solar rotation. The apparent merging of the northern and middle lobes may be due to a rotation effect.
In the meantime, the southern lobe evolves but does not interact with the two other EUV lobes.

\subsection{Magnetic configuration from PFSS extrapolations}\label{ss:PFSS_PS}

We now use Potential Field Source Surface (PFSS) extrapolations \citep{altschulerMagneticFieldsStructure1969, stansbyPfsspyPythonPackage2020} in order to examine the large-scale magnetic topology of this event. We use the Air Force Data Assimilative Photospheric Flux Transport \citep[ADAPT;][]{argeAirForceData2010, argeImprovingDataDrivers2011, argeModelingCoronaSolar2013, hickmannDataAssimilationADAPT2015} full-Sun modeled magnetograms, based on observed magnetograms from the Helioseismic and Magnetic Imager \citep[HMI;][]{scherrerHelioseismicMagneticImager2012} onboard SDO (hereafter refereed as ADAPT-HMI). ADAPT is a flux-transport model, which starts with an ensemble of realizations and evolves them using flux transport processes such as differential rotation, meridional circulation and supergranulation \citep{wordenEvolvingSynopticMagnetic2000}. Observations are assimilated in the model accounting for model and observational uncertainties. We refer the reader to the ADAPT papers for more details on these data products.

\begin{figure} 
\centerline{\includegraphics[width=\textwidth,trim=250 250 150 250,clip]{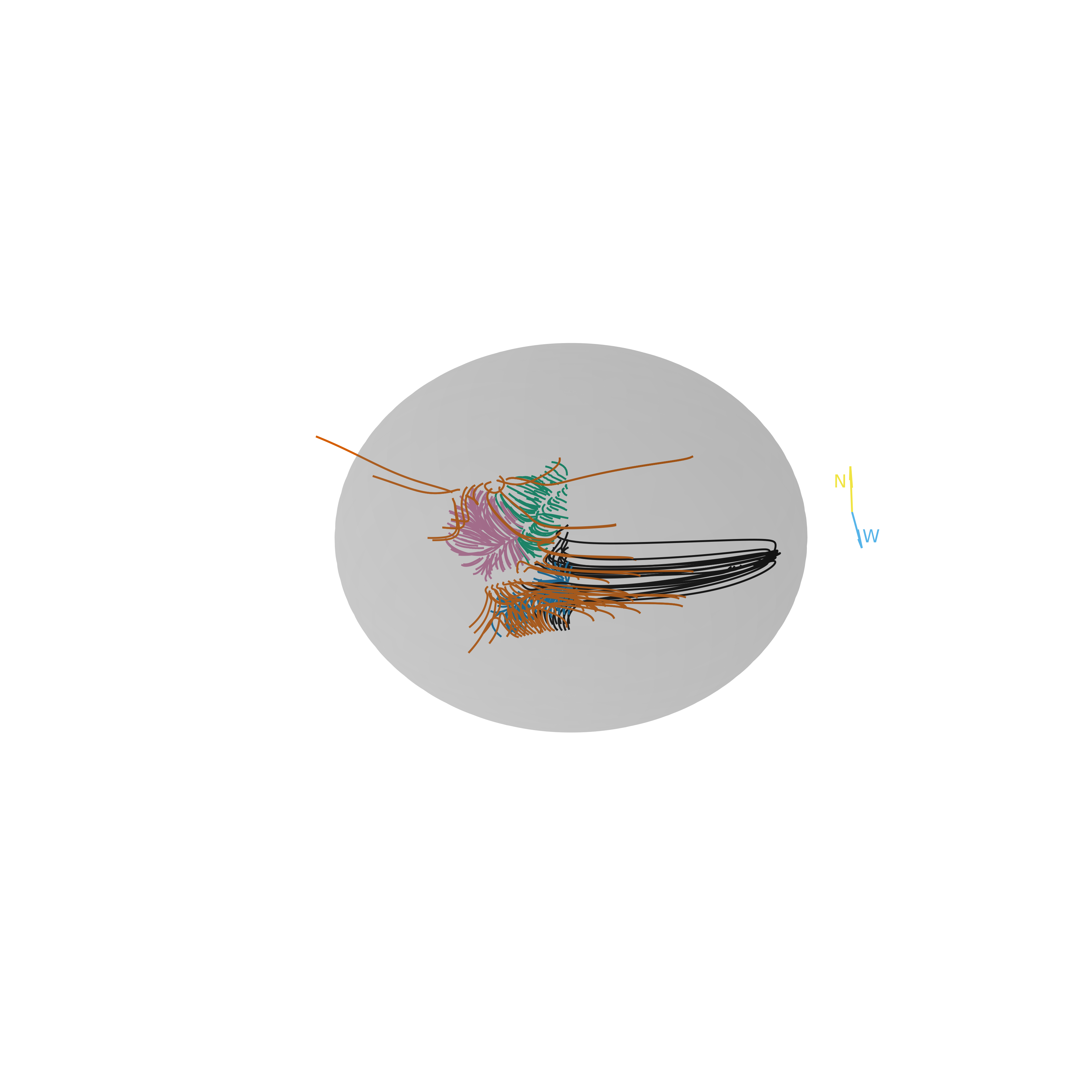}}
\caption{The two pseudo-streamer topologies from the PFSS extrapolation using the \hbox{ADAPT-HMI} magnetogram of June 14, 2012 08:00 UT. In this 3D view, the Sun is rotated in order to observe the topologies from above. The north and west directions are indicated in yellow and light blue, respectively. Open magnetic field lines, all of positive polarity, are represented in orange. For the northern pseudo-streamer, closed field lines marking the PS-dome are represented in green for the western section that is oriented north-west and in pink for the eastern section that is more inclined toward the east-west direction. The closed field lines of the southern pseudo-streamer are represented in dark blue. Finally, the large closed field lines that are bounding the two pseudo-streamers are represented in black.}\label{fig:fig2_topology}
\end{figure}

On Figure~\ref{fig:fig2_topology}, we show the result of the extrapolation for June 14, 2012 08:00 UT in the area corresponding to the ROI only.
We trace field lines on a near-uniform spherical grid spanning from (\SI{-65}{\degree},\SI{-20}{\degree}) in Carrington latitude and (\SI{340}{\degree},\SI{10}{\degree}) in Carrington longitudes, with a resolution of about \SI{2}{\degree} both in longitude and in latitude. The source surface is fixed at 2~R$_{\odot}$. We note that the choice of this height does not impact our results. 

Instead of having  three pseudo-streamer topologies as expected from the EUV structures, the PFSS extrapolation shows one pseudo-streamer topology in the north (green and pink closed field lines bounded by orange open field lines and large black closed field lines) and a second pseudo-streamer topology in the south (blue closed field lines bounded by orange open field lines and large black closed field lines). Contrary to a null-point topology that consists of one null-point embedded in a unipolar field, the magnetic topology for a pseudo-streamer is more complex. The underlying magnetic structure of a pseudo-streamer consists of a closed field intersecting a uni-polar open field. The basic topological elements are a dome separatrix surface (hereafter called the PS-dome) confining small closed loops, an open vertical fan or separatrix curtain that intersects the dome and which is located at the interface between two open-field regions of the same polarity (positive polarity, orange field lines in the Figure). The field surrounding the PS-dome is a combination of open and closed field bounding the PS-dome \citep[see][for details]{titovMagneticTopologyCoronal2011, titov2010August122012}.
The northern pseudo-streamer topology covers the same latitudes as the northern and middle lobes described in Figure~\ref{fig:fig1_context} (from \SI{-20}{\degree} to \SI{-45}{\degree} in Carrington latitude). Its orientation changes with the longitude. The western section of the PS-dome shown by green field lines on Figure~\ref{fig:fig2_topology} is oriented along the north-west direction, while the eastern section shown by pink field lines is leaning toward the east-west direction. The southern pseudo-streamer consists of a closed flux domain (dark blue field lines on Figure~\ref{fig:fig2_topology}) embedded in open field or large closed loops.

The PFSS model allows us to determine the  topology of the magnetic field, but it does not capture its precise configuration. First, the model assumes a potential field, which inherently excludes electric currents. However, the solar corona is characterized by impulsive events occurring at all scales, where such currents are present.
Second, the PFSS extrapolation relies on a ADAPT-HMI map in which the region is located outside of the co-temporal HMI disk observations approximately in the range \SI{20}{\degree} to \SI{170}{\degree} in Carrington longitude. 
However, ADAPT-HMI maps, unlike synoptic maps, are not relying on old photospheric magnetic field measurements.
Instead, the far-side areas are filled in thanks to helioseismological data estimating the presence of active regions and the flux-transport model \citep[see more details in][]{hickmannDataAssimilationADAPT2015}. Small-scale activity (magnetic flux emergence, coronal activity, etc), may still have altered the magnetic field distribution. Nonetheless, the large-scale photospheric field is still expected to produce a similar global coronal magnetic topology.
We checked this by using ADAPT-HMI magnetograms on June 22, 2012, i.e. after the event when our ROI is supposedly located around the prime meridian (not shown here). The global magnetic field topology: the two pseudo-streamer lobes embedded in open field, are similar. Differences appear at smaller scales, which we are not aiming at commenting.
Despite these caveats, we can thus confront the EUV structures with the magnetic field topology and identify the magnetic field associated with the evolution of the EUV structures. 

\begin{figure} 
\centerline{\includegraphics[width=0.8\textwidth]{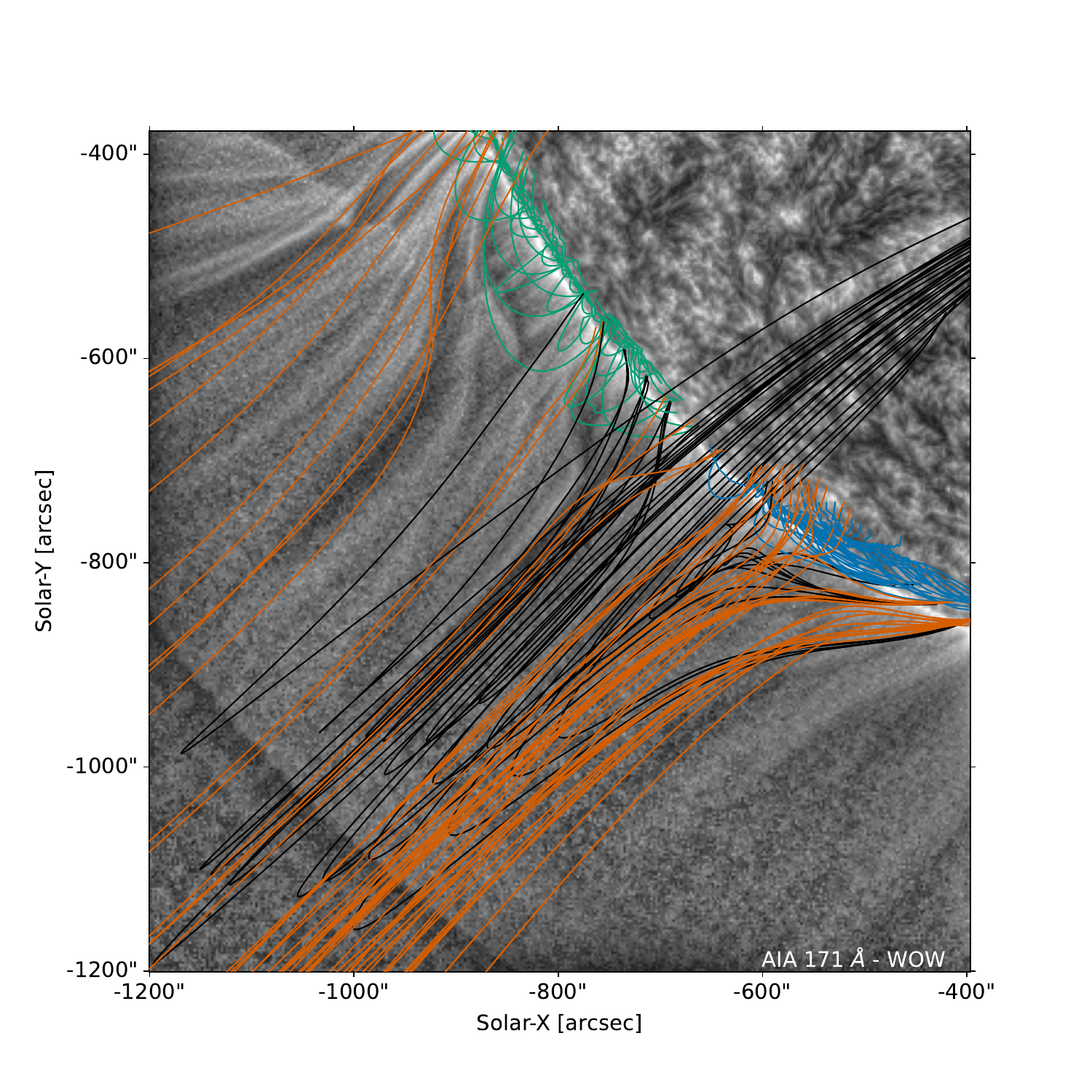}}
\caption{Magnetic field lines from the PFSS extrapolation presented in Figure~\ref{fig:fig2_topology}, which are superimposed to the \SI{171}{\angstrom} image on June 14, 2012 00:00 UT, processed with the WOW algorithm (first snapshot of Figure~\ref{fig:fig1_context}). The colors remain the same for the different groups of field lines as in Figure~\ref{fig:fig2_topology}. Pink field lines are not traced as most of their footpoint would erroneously appear on the disk even though they are still at the far side at the beginning of the event.}
\label{fig:fig3_PS_EUV_comparison}
\end{figure}

\begin{figure} 
	\resizebox{\hsize}{!}
	{\begin{tabular}{cc} 
		\centerline{\includegraphics[width=\textwidth,trim=100 100 150 250,clip]{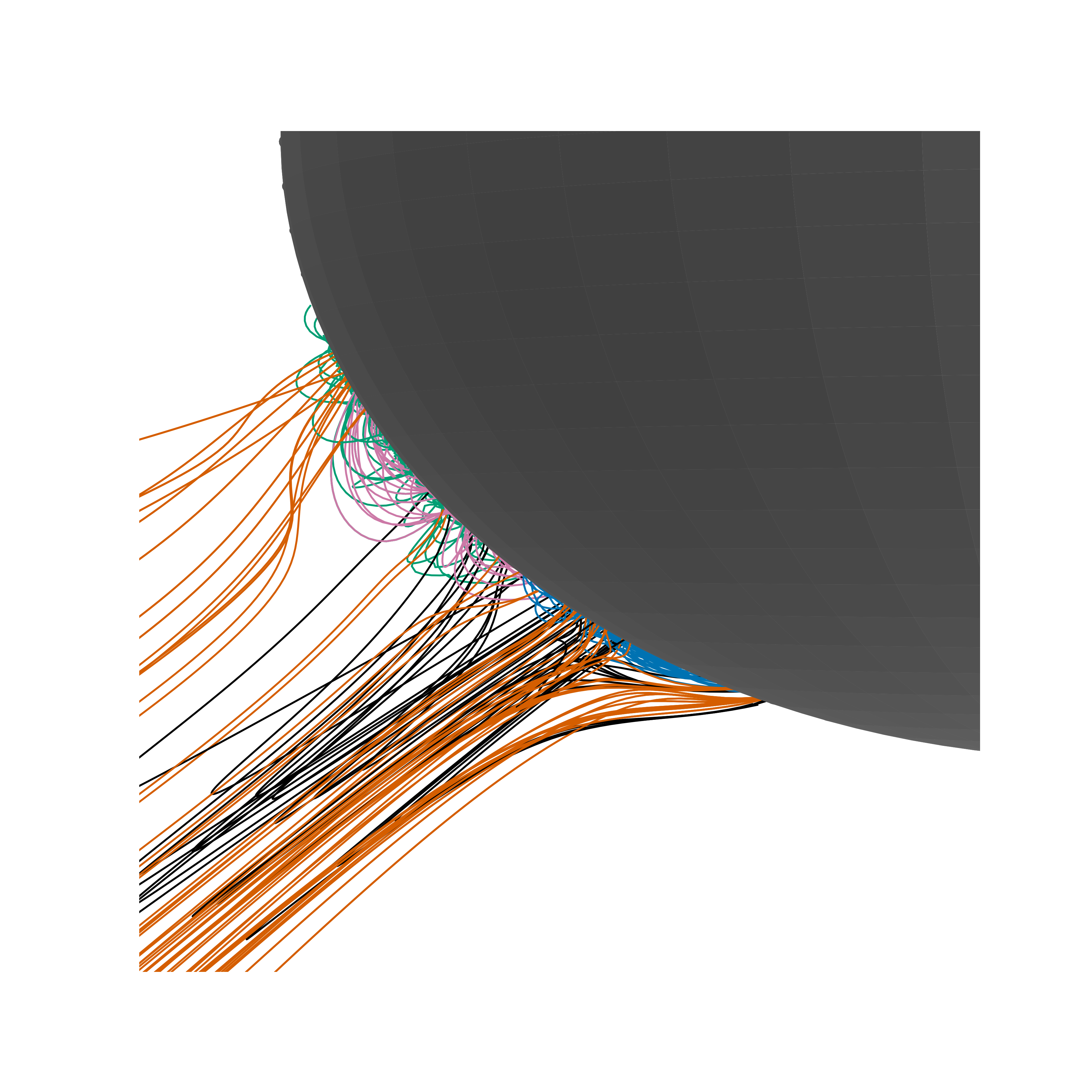}} & \centerline{\includegraphics[width=\textwidth,trim=100 100 150 250,clip]{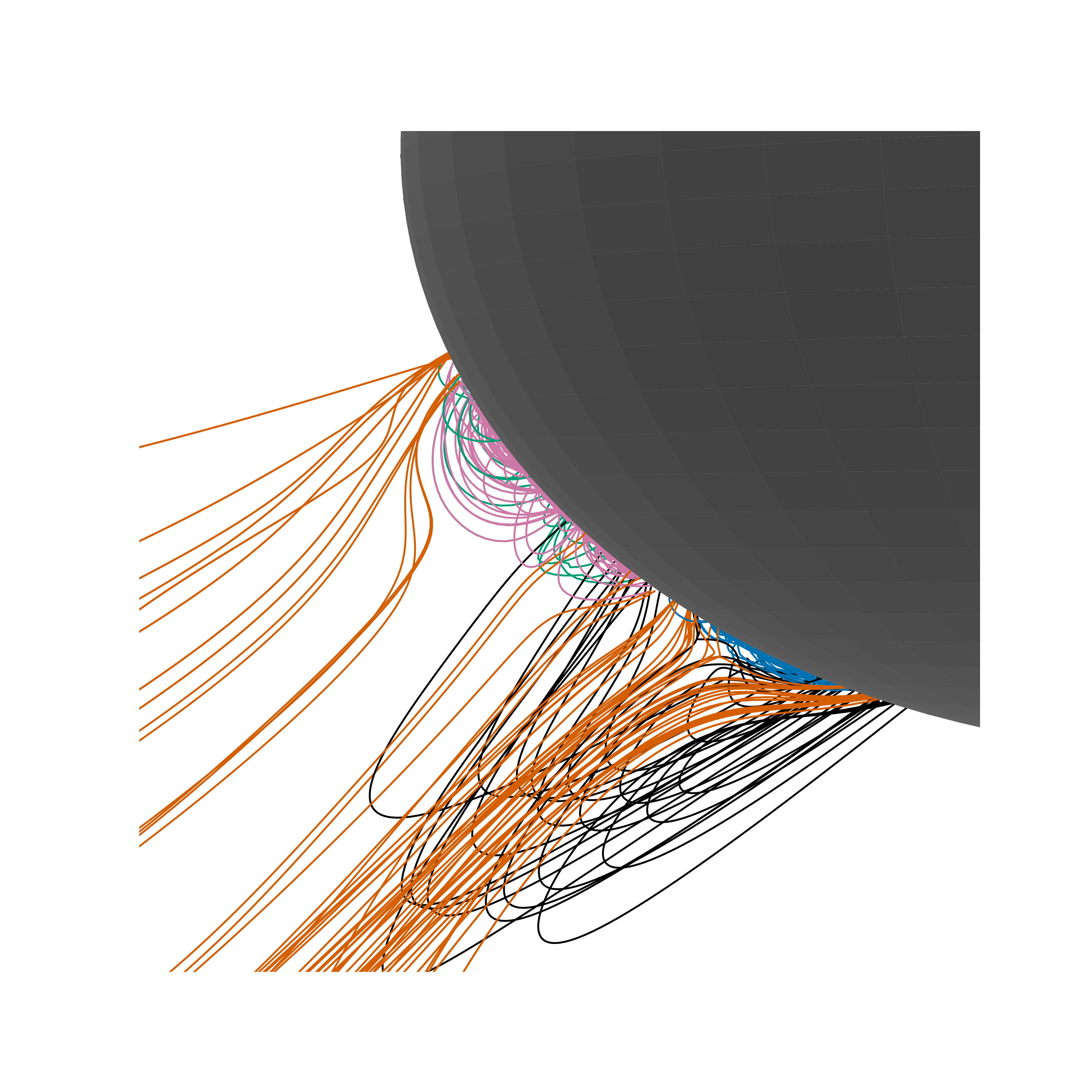}}
        
	\end{tabular}
	}
\caption{Rotation of the 3D PFSS view with all the different groups of field lines (same colors as in Figure~\ref{fig:fig2_topology} and Figure~\ref{fig:fig3_PS_EUV_comparison}). Left panel: Earth view at the beginning of the event. Right panel: Rotation of \SI{14}{degrees} in longitude mimicking about a day of solar rotation.}
\label{fig:fig4_PS_rotation_view}
\end{figure}

\subsection{Comparison of the pseudo-steamers evolution from the PFSS and EUV emission}\label{ss:comparison_PFSS_EUV}

In Figure~\ref{fig:fig3_PS_EUV_comparison}, we overlay field lines selected from the PFSS extrapolation presented in Figure~\ref{fig:fig2_topology} on the \SI{171}{\angstrom} image on June 14, 2012 00:00 UT, processed with the WOW algorithm (first snapshot in Figure~\ref{fig:fig1_context}).
Contrary to what we hinted from the EUV observations (Section~\ref{ss:euv_PS_observation}) the northern and the middle EUV lobes are not separated structures divided by open magnetic flux. Instead, they are associated with the PS-dome, i.e. western (green lines) and eastern (pink lines) small closed field lines, respectively, and open field lines of the northern pseudo-steamer topology.  
We note that some of the field lines on Figure~\ref{fig:fig3_PS_EUV_comparison}, that seem to appear on disk, are actually coming from the far side. For this reason we do not overlay the pink field lines on Figure~\ref{fig:fig3_PS_EUV_comparison}. Indeed, most of their footpoints are at longitudes below \SI{350}{\degree} which would confuse the picture. Instead, we plot them with all the other groups of lines in Figure~\ref{fig:fig4_PS_rotation_view} in order to study the change of appearance of the structures caused by the solar rotation and its impact on the long-term evolution of the northern and middle EUV lobes. In this Figure, the left panel shows an Earth view of the pseudo-streamers topologies at the beginning of the event, while the right panel shows a rotated view by about \SI{14}{\degree} in longitude which is about the rotation rate per day. At the beginning of the event, the curved geometry of the closed flux below the PS-dome implies that from the Earth line-of-sight (LOS) the western PS-dome section (green lines) is located northward while the eastern PS-dome section (pink lines) is located southward. With time and the solar rotation, the green field lines from the western section change their orientation with respect to the LOS.
Meanwhile, the pink field lines from the eastern PS-dome section are more and more visible and superimposed to the green lines. We also notice that the open field lines and large closed field lines bounding the eastern PS-dome section in the south intersect the LOS at the beginning of the event and then progressively move to the south. Thus, the northern and the middle lobes are in reality from the same PS-dome but with a change in its orientation. This reconciles the pseudo-streamer topology and the EUV observations.

Concerning the southern EUV lobe, it is associated with the southern pseudo-streamer. The EUV emission seems to be mostly linked to the large closed field lines (in black) and the open field lines. The dark blue closed lines appear too low lying to largely contribute to its dynamics in altitude even though they may transiently be visible as in the snapshot 3 and 4 of Figure~\ref{fig:fig1_context}.  Moreover, this southern lobe evolves differently than the two other lobes. In particular, the dynamics of this lobe seems to be caused by several small-scale eruptions and/or solar jets (see the animation associated with Figure~\ref{fig:fig10_TNE_open_structure}). Our study focuses on non-eruptive pseudo-streamer dynamics. 
This southern lobe is therefore not used in this work, we will not discuss it further.

\section{Signatures of thermal non-equilibrium (TNE) in the pseudo-streamer}\label{s:TNE_signatures}

\subsection{Long-period EUV pulsations}\label{ss:pulsations_analysis}

\begin{figure} 
\centerline{\includegraphics[width=\textwidth,clip=]{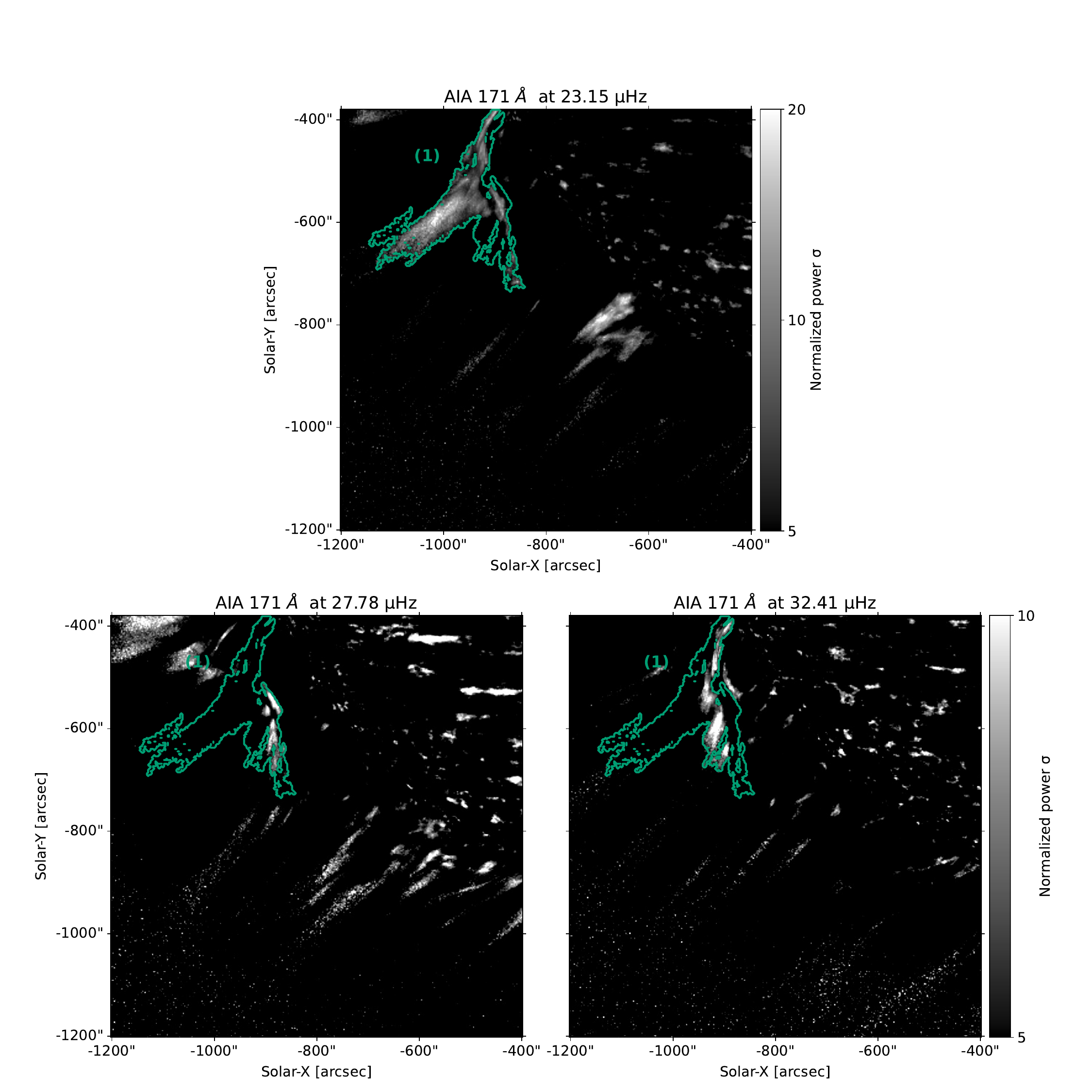}}
\caption{Detection of long-period EUV pulsations in the northern and middle EUV lobes (northern pseudo-streamer topology). Top panel: Normalized power maps for the \SI{171}{\angstrom} AIA channel at the dominant frequency \SI{23.15}{\mu \hertz} which corresponds to 12 hours of period. 
Bottom: Complementary view of the detection at \SI{27.78}{\mu \hertz} and \SI{32.41}{\mu \hertz}, which are the two immediately succeeding frequency bins.
The detected contour (1) is shown in green. The logarithmic scale is saturated at $20\sigma$ for the top panel and at $10\sigma$  for the bottom panels.}
\label{fig:fig5_power_maps}
\end{figure}

\subsubsection{Method of detection}
In the ROI, we identify several regions exhibiting long-period EUV pulsations. These are detected 
using the algorithm of \citet{auchereLongPeriodIntensityPulsations2014} on the unfiltered "pulsation dataset". This is the same procedure that was used in previous studies \citep{fromentEvidenceEvaporationincompleteCondensation2015, auchereCoronalMonsoonThermal2018, fromentMultiscaleObservationsThermal2020, pelouzeSpectroscopicDetectionCoronal2020, sahinSpatialTemporalAnalysis2023}. With this method we analyze the ROI pixel-per-pixel and are thus not looking for pulsations in specific underlying coronal features. We resample the time series using a linear interpolation to ensure a regular cadence. The resampling does not affect the detection of long-period pulsations in the Fourier analysis, as it may only affect higher frequencies. In order to increase the signal-to-noise ratio (S/N), we further bin the images $4\times4$. We compute a power spectra cube by performing a Fourier transform of the images cube along the time axis. The spectra are normalized to the variance of the light curves. The Fourier power (hereafter normalized power) is expressed as excess above an estimation of the average fluctuation level $\sigma(\nu)$, that depends on the frequency and is computed for each spectrum.
In order to determine whether an excess of power above this background is significant or not, we use the global probability \citep{scargleStudiesAstronomicalTime1982, gabrielSearchSolarModes2002, auchereFourierWaveletAnalysis2016} that at least one frequency bin among N/2 shows an excess of power m times greater than $\sigma(\nu)$, which is expressed as follows: $P_g(m) = 1-(1-e^{-m})^{N/2}$, with N the number of time steps for each time series. An excess of power of $m=10\sigma$ then corresponds to a 1~\% chance that the signal is due to random fluctuations i.e., we have at least a 99~\% confidence level on our detections. However, for a large ROI such as the one studied in this paper, it is likely that  random isolated pixels may show a significant excess of power. We therefore look for spatial coherence by clustering 6-connected voxels, adjacent on one of their faces along the spatial or the frequency direction, that have excess of power above the $10\sigma$ threshold. Finally, we discard the regions that are smaller than an arbitrary threshold of 150 pixels.
The area of the remaining selected regions, i.e. detections, are then extended to the adjacent $5\sigma$ voxels as long as the confidence level of the detection on light curve averaged the region remains $10\sigma$. We end up with detected contours automatically selected from the normalized power maps.
As for the other events studied off-limb \citep{auchereCoronalMonsoonThermal2018, fromentMultiscaleObservationsThermal2020, sahinSpatialTemporalAnalysis2023}, we turned-off the differential rotation compensation feature of the algorithm. Even though variations in height of coronal structures are thought to be limited off-limb during 2.5-day sequences, we discussed in Section~\ref{s:PS_context} about the effect of the solar rotation on the appearance of the studied structure. The detected contours of long-period intensity pulsations may thus show some degree of geometric distortion (see Section~\ref{ss:comparison_PFSS_EUV}). Finally, we only consider frequencies between \SI{187}{\micro\hertz} and \SI{18}{\micro\hertz} in order to ensure at least four periods in the time series.

\subsubsection{Main event detected}
We detect a main region with long-period EUV pulsations in the ROI, that is located in the area structured by the northern and middle EUV lobes, i.e. the northern pseudo-streamer. In the first panel of Figure~\ref{fig:fig5_power_maps} we show the normalized power map centered at \SI{23.15}{\micro\hertz} (i.e. 12 hours), computed for the \SI{171}{\angstrom} channel which is the passband for which we found the strongest signals. Contour (1) encloses the detection, i.e. where the power is spatially coherent above $10\sigma$ at this dominant frequency. As previously explained, the final detected contours, which contain the detected regions, may not always match the power map in one frequency bin, since it may correspond to a detection over several frequency bins. This is the case for contour (1). The spatial extension of this contour is better understood by looking additionally at the normalized power map centered at \SI{27.78}{\micro\hertz} (i.e. 10 hours) and \SI{32.41}{\micro\hertz} (i.e. 8.6 hours). For example, in the extension of contour (1) located in the box defined by $\rm{Solar-X} =[\SI{-945}{\arcsecond};\SI{-900}{\arcsecond}]$ and $\rm{Solar-Y} =[\SI{-600}{\arcsecond};\SI{-650}{\arcsecond}]$, there is only significant power in the \SI{32.41}{\micro\hertz} frequency band.

Southward of the area covered by contour (1), we notice a second large region of significant power in the map at \SI{23.15}{\micro\hertz} in the box defined by $\rm{Solar-Y} =[\SI{-700}{\arcsecond};\SI{-900}{\arcsecond}]$ and $\rm{Solar-X} =[\SI{-780}{\arcsecond};\SI{-580}{\arcsecond}]$. A contour (not shown here) is also detected automatically in this region. However, we choose here not to analyze this event since it is located in the southern EUV lobe, i.e. southern eruptive pseudo-streamer that is not relevant for our study (see Section~\ref{ss:comparison_PFSS_EUV}).

\begin{figure} 
\centerline{\includegraphics[width=\textwidth,clip=]{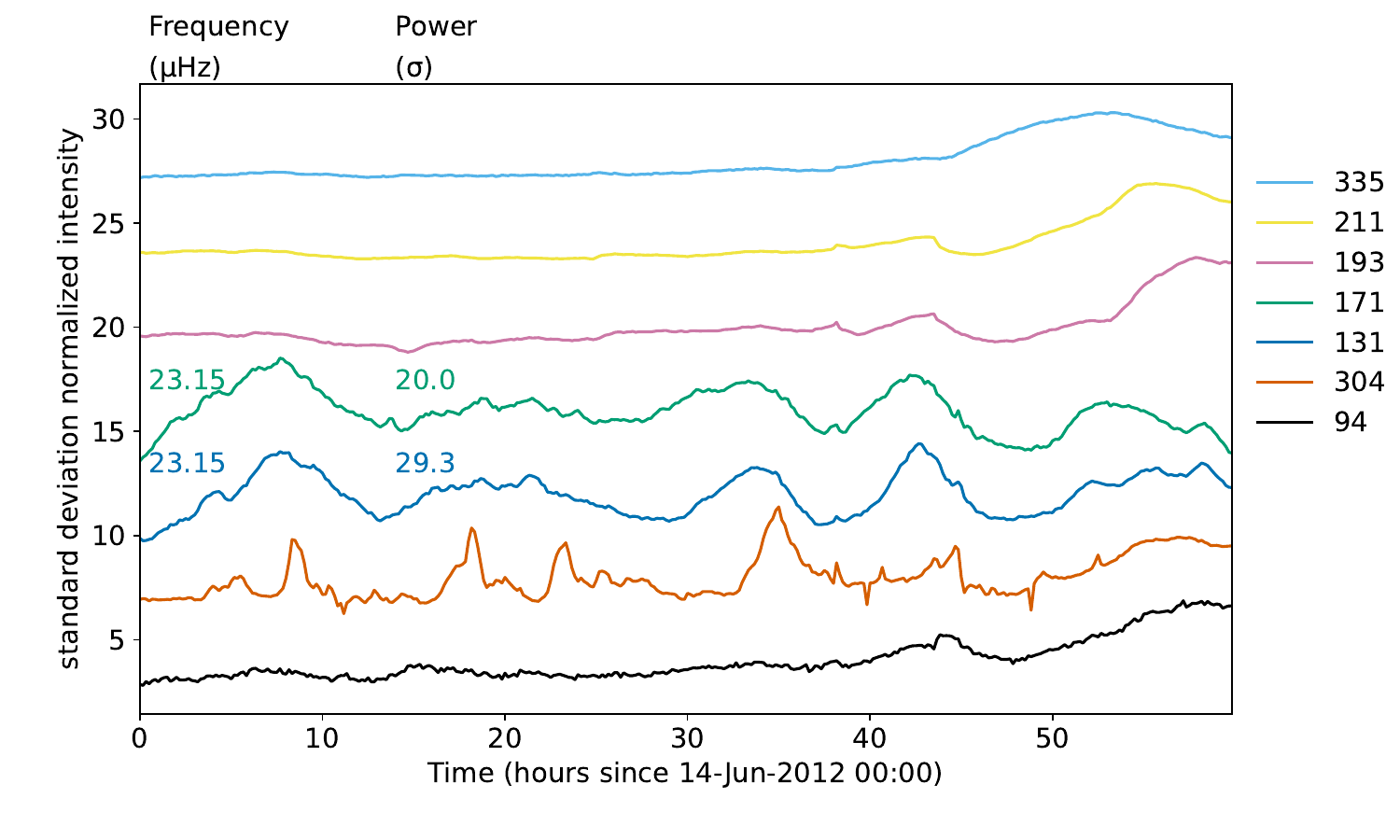}}
\caption{Light curves, averaged over the area enclosed by the detected contour (1), in the seven AIA channels studied. For each time series, the mean value is subtracted, and the result is then normalized by the standard deviation. They are then arbitrarily shifted by 4 on the y-axis. When a periodicity is detected, we display the corresponding frequency and normalized power.}
\label{fig:fig6_light_curves}
\end{figure}

In Figure~\ref{fig:fig6_light_curves} we show the light curves, averaged over the area enclosed by contour (1) for the seven selected AIA channels.
We detect a dominant periodicity of 12 hours (i.e., \SI{23.15}{\micro\hertz}) in the \SI{171}{\angstrom} and \SI{131}{\angstrom} channels with an excess of power of $20.0\sigma$ and $29.3\sigma$, respectively. This corresponds to a probability of random occurrence of $10^{-7}$ and $10^{-11}$, respectively. Even though no periodicity is detected in the \SI{304}{\angstrom} channel, some synchronicity can be noticed. Indeed, recurrent \SI{304}{\angstrom} pulses are following the modulation in the \SI{171}{\angstrom} and \SI{131}{\angstrom} channels.
In Section~\ref{ss:rain_analysis} we will dive into the analysis of the \SI{304}{\angstrom} channel images and light curves, in particular to determine the link between the light curves variability and coronal rain events.

There is no detected periodicity in the \SI{335}{\angstrom}, \SI{211}{\angstrom}, \SI{193}{\angstrom} and \SI{94}{\angstrom} channels and no related peaks, except for the pulse located about 48 hours after the start of the sequence. After carefully examining the  EUV images in these bands, it is clear that the EUV structures (lobes) linked to the pseudo-streamers topologies are not seen in these channels. This could be due to the temperature structure of the pseudo-streamer, which does not reach the peak temperature response of these channels (above 1~MK), similarly as in \citet{guennouLifecycleLargescalePolar2016}. In the rest of the analysis, we will thus no longer include these channels.

\subsubsection{Additional events detected} 

\begin{figure} 
\centerline{\includegraphics[width=\textwidth,clip=]{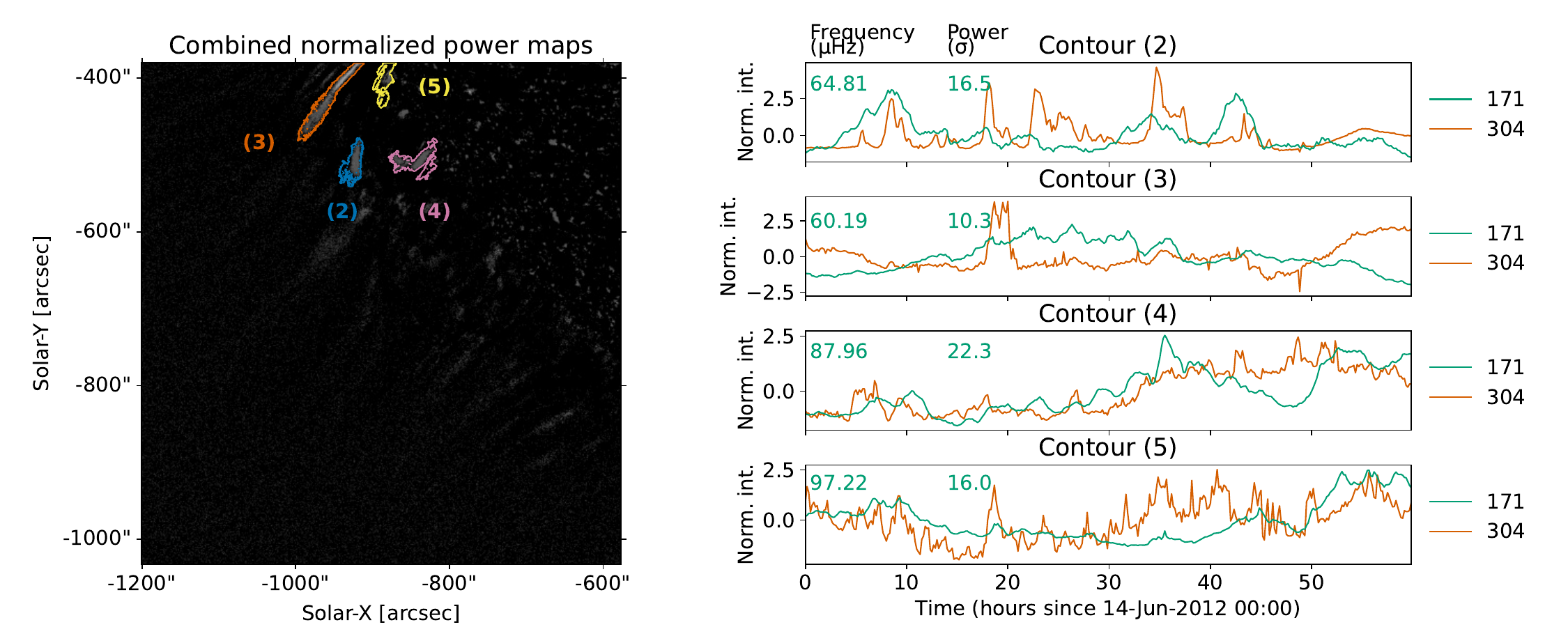}}
\caption{Additional long-period EUV pulsations events detected using the non-spatially binned "pulsation dataset". Right: Contours enclosing the detections on a combination of the normalized power maps (power above 8$\sigma$ at the dominant frequency of each detection : contour (2) at \SI{61.8}{\mu \hertz} (i.e. 4.5 hours), contour (3) at \SI{60.19}{\mu \hertz} (i.e. 4.6 hours), contour (4) at \SI{87.96}{\mu \hertz} (i.e. 3.2 hours) and contour (5) at \SI{97.22}{\mu \hertz} (i.e. 2.9 hours). Left: Light curves, averaged over the detected contours, in the \SI{171}{\angstrom} and \SI{304}{\angstrom} channels only for conciseness. For each time series, the mean value is subtracted, and the result is then normalized by the standard deviation. The frequency detected is displayed along with corresponding normalized power.}
\label{fig:fig7_additionnal_contours}
\end{figure}

Our criteria for detecting long-period EUV pulsations are stringent. We do not aim to present exhaustive detections in the ROI explored. 
Moreover, binning our images $4\times4$ effectively increases the S/N and yields unified detection contours, avoiding the fragmentation observed in non-spatially binned data. It ensures that our analysis focuses on the most robust and significant regions. 
However, the combination of this binning and the minimal size threshold for final regions may filter out some smaller regions. We thus re-ran the detection algorithm on the non-spatially binned "pulsation dataset". We detect four additional long-period EUV pulsations events in the \SI{171}{\angstrom} channel. These are located the northern pseudo-streamer. For conciseness,  we overlay all of them in Figure~\ref{fig:fig7_additionnal_contours} on the combined normalized power maps at the dominant frequency of each detection. Contour (2) is detected at \SI{64.81}{\mu \hertz} (i.e. 4.3 hours) with an excess of power of $16.5\sigma$, contour (3) at \SI{60.19}{\mu \hertz} (i.e. 4.6 hours) with an excess of power of $10.3\sigma$, contour (4) at \SI{87.96}{\mu \hertz} (i.e. 3.2 hours) with an excess of power of $22.3\sigma$ and contour (5) at \SI{97.22}{\mu \hertz} (i.e. 2.9 hours) with an excess of power of $16\sigma$.
No periodicities are detected in the \SI{304}{\angstrom} channel, and for some of the time series (contours (3,4,5)) the time series seem un-correlated with the \SI{171}{\angstrom} ones. Conversely, the \SI{131}{\angstrom} channel time series (not shown here) are very similar to the \SI{171}{\angstrom} ones.
While contour (1) was encompassing a large portion of the northern pseudo-streamer, these additional contours are linked to smaller sub-structures. Contour(2) is embedded within contour (1), which explain their similar associated time series. Without further commenting on the location of these contours now (see Section~\ref{ss:rain_topo}),  we would like to mention that for such small contours, projection effects likely become important. Indeed, the dataset analyzed spans 2.5 days. The contours constructed in the Fourier space are fixed in time and when we construct the final time series, averaged over the area delimited by these contours, the pseudo-streamer has rotated. This may have a strong effect in the area dominated by the small loops under the PS-dome as explained in Section~\ref{ss:comparison_PFSS_EUV}.

\subsection{Recurring coronal rain showers}\label{ss:rain_analysis}

We now proceed to study the properties of the EUV pulsations. In particular, we study the periodic cooling of these regions. We proceed similarly as in previous studies of long-period EUV pulsations, in the case of coronal loops \citep[e.g.,][]{auchereCoronalMonsoonThermal2018, fromentMultiscaleObservationsThermal2020}:  (1) by studying the coronal rain showers timings and locations (this Section), (2) by studying the sequential appearances of the pulses in different channels according to their peak temperature response (Section~\ref{ss:cooling_analysis}).

\begin{figure} 
\centerline{\includegraphics[width=\textwidth,clip=]{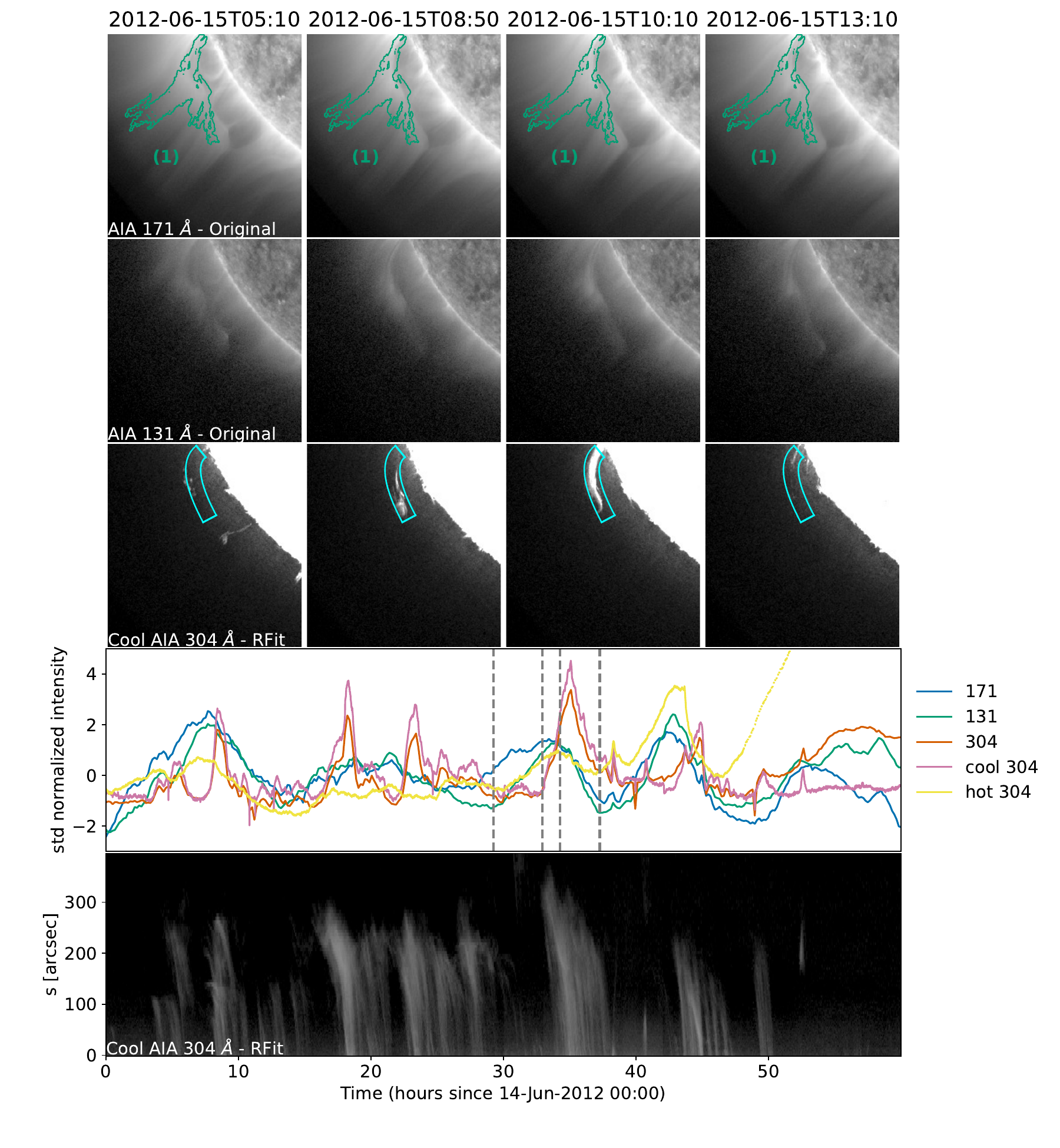}}
\caption{Periodic cooling signatures in the northern pseudo-streamer. First three rows: Key stages observed during the third EUV pulse (between 30 and 40 hours after the start of the sequence). The times of these snapshots are marked sequentially in the time series below (fourth row), with gray vertical dashed lines. The first row shows the \SI{171}{\angstrom} channel sequence (unfiltered, in log scale) on which we overlay contour (1) in green. It is not overlaid on the images of the other channels for visibility reasons. In the second row, we display the \SI{131}{\angstrom} channel sequence (unfiltered, in log scale). The third row shows the cool \SI{304}{\angstrom} emission extracted with the RFit method. Fourth row: Light curves averaged over the region enclosed by contour (1) in the \SI{171}{\angstrom}, \SI{131}{\angstrom}, and \SI{304}{\angstrom} channels. The hot and cool \SI{304}{\angstrom} components time series are also shown. 
Fifth row: Time-distance plot showing the coronal rain showers. The cool \SI{304}{\angstrom} intensity is averaged over the width of the slits which are roughly perpendicular to the rain path. These slits form the cyan contour in the third-row images. The origin of the y-axes is the northern part of the cyan contour. An animation is available for this figure. We note that while the images are from the binned "pulsation dataset" for an easy time synchronization, the \SI{304}{\angstrom} times series and time-distance map are from the "rain dataset".}
\label{fig:fig8_TNE_overview}
\end{figure}

In the \SI{304}{\angstrom} "rain dataset" sequence of images (see the accompanying movie of Figure~\ref{fig:fig_11_rain_types}), we see recurring and transient coronal rain showers, forming in the different parts of the northern pseudo-streamer and falling toward the surface. These showers appear most prominently within contour (1), and with a time delay compared to the five main quasi-synchronous EUV pulses, as counted in the light curve averaged over contour (1) in the \SI{171}{\angstrom} and \SI{131}{\angstrom} AIA channels.
Here, we will only comment on the appearance of the rain related to contour (1),  which is the only location where the showers are recurring. Our results will be also applicable to contour (2), since it is embedded within contour (1). In Section~\ref{ss:rain_dynamic}, we will widen our analysis to cover the transient coronal rain showers that are related to all the other contours and to other parts of the northern pseudo-streamer.

In order to better identify the coronal rain showers in the \SI{304}{\angstrom} images and light curves, we extract the cool and hot components of the \SI{304}{\angstrom} channel by using the Response fit (RFit) method introduced in \citet{antolinDecomposingAIA304Channel2024}. The method is as follows. The \SI{304}{\angstrom} response function has two main components: one at about 0.08~MK, due to the He II \SI{303.8}{\angstrom} emission that dominates the channel, and one at about 1.8~MK. This later hot coronal component is well covered by all the other EUV channels. This means that it can be modeled by a linear combination of the other bands. The obtained hot \SI{304}{\angstrom} intensities are then subtracted from the original \SI{304}{\angstrom} intensities in order to obtain the cool \SI{304}{\angstrom} ones. We apply this method to the "rain dataset" (full spatial resolution and \SI{1}{\minute}  of cadence) in order to fully capture the relevant coronal rain dynamics.
We group images from the seven passbands into septuplets for application of the RFit method. Times at which at least one channel has an exposure time below \SI{0.5}{\second} are discarded. This corresponds to 68 images in the 2.5-days time sequence.

The cool \SI{304}{\angstrom} images will be used to study the rain while the hot \SI{304}{\angstrom} intensities will later complement the analysis of the cooling in the light curves, i.e. the study of the sequential appearances of the EUV pulses in different channels according to their peak temperature response (see Section~\ref{ss:cooling_analysis}).

In Figure~\ref{fig:fig8_TNE_overview}, we show again the light curves averaged over the region enclosed by contour (1) in the \SI{171}{\angstrom}, \SI{131}{\angstrom}, and \SI{304}{\angstrom} channels. 
This time the \SI{304}{\angstrom} light curve is from the "rain dataset", hence the very small differences compared to the curve shown in Figure~\ref{fig:fig6_light_curves} ("pulsation dataset"). We now also include the cool and hot \SI{304}{\angstrom} light curves. We observe that the \SI{304}{\angstrom} peaks are dominated by the cool \SI{304}{\angstrom} component, except in the case of the fifth EUV pulse (as counted in the \SI{171}{\angstrom} channel), for which the hot \SI{304}{\angstrom} component dominates. We note that the hot \SI{304}{\angstrom} time series is normalized using the data from the first 50 hours of the sequence, that is why the last few hours are displayed in dotted line.

In the top panels of Figure~\ref{fig:fig8_TNE_overview}, we highlight four key stages (\SI{171}{\angstrom}, \SI{131}{\angstrom}, and cool \SI{304}{\angstrom} images) observed during the third EUV pulse (as counted in the \SI{171}{\angstrom} channel), i.e. between 30 and 38 h after the start of the sequence. In this sequence of images, we overlay contour (1) in green as well as a cyan contour that is encapsulating the main location of coronal rain showers seen within contour (1). The coronal rain showers as well as the intensity variations in the sequence of images can be tracked using the animation accompanying Figure~\ref{fig:fig8_TNE_overview}.

On snapshot 1 (left column of Figure~\ref{fig:fig8_TNE_overview}, on June 15, 06:00 UT), the intensity is starting to increase in the \SI{171}{\angstrom} channel and is still at its minimum in the \SI{131}{\angstrom} and \SI{304}{\angstrom} channels. We notice some faint coronal rain clumps appearing, which explain the small peak beginning to rise in the \SI{304}{\angstrom} channel. On snapshot 2 (June 15, 09:40 UT), the intensity is at its maximum in the \SI{171}{\angstrom} and \SI{131}{\angstrom} channels and coronal rain is starting to frankly appear in contour (1).  On snapshot 3 (June 15, 11:00 UT), while the intensity in the \SI{171}{\angstrom} and \SI{131}{\angstrom} channels starts to slightly decrease, the coronal rain shower is at its peak. The rain path forms a hook-like feature. On snapshot 4 (June 15, 14:00 UT), the coronal rain event has ended and we are back at a minimum of intensity for all the channels.

To obtain a general view of the full coronal rain evolution, we construct a time-distance map of the averaged cool \SI{304}{\angstrom} intensity ("rain dataset"). This map is based on slits roughly perpendicular to the rain path, within the cyan contour shown on the third row of Figure~\ref{fig:fig9_time-lag_maps}. By comparing the time-distance evolution of the rain showers (fifth row of Figure~\ref{fig:fig9_time-lag_maps}) and the light curves, we can see that the cool \SI{304}{\angstrom} peaks correspond to the coronal rain showers. These showers are recurring and follow with the main EUV pulses in the \SI{171}{\angstrom} and \SI{131}{\angstrom} AIA channels.
However, for the fifth EUV pulse, there is no more rain in contour (1) and the pseudo-streamer has rotated enough so only the southern tip of the western part of the contour is within the EUV lobe. The transient signal at about 52 hours since the start of the sequence is coming from an unrelated solar jet entering contour (1) and the cyan contour.
We also notice that there are sometimes two (or more) main peaks in the \SI{304}{\angstrom} channel corresponding to one single EUV pulse in the \SI{171}{\angstrom} and \SI{131}{\angstrom} channels. This is the case for the two first EUV pulses covering from the beginning to 30 hours into the sequence. These shorter timescales may be linked to the shorter periods (4.3~hours) highlighted by the detection of EUV pulsations in contour (2) or to distinct rain shower events due to LOS superimposition. This rain dynamics will be described in Section~\ref{ss:rain_dynamic}.

\subsection{Cooling pattern in the light curves}\label{ss:cooling_analysis}

\begin{figure}
\centerline{\includegraphics[width=\textwidth]{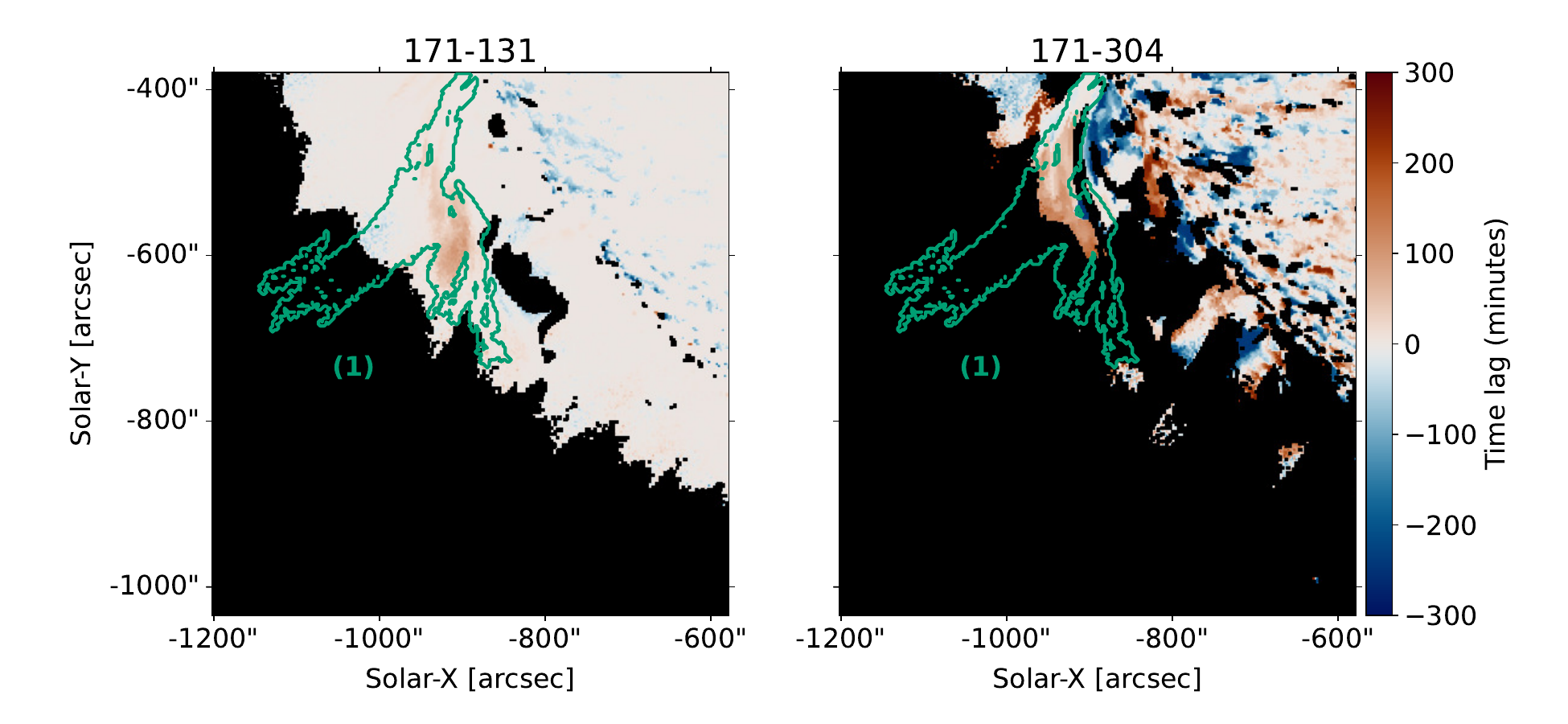}}
\caption{Time-lag maps with peak cross-correlation values in the 171–131 and 171-304 pairs of channels using the time series between 26 and 50 hours after the start of the sequence i.e. covering only the EUV pulses 3 and 4.  Masked (black) areas correspond to peak cross-correlation values under 0.2. Contour (1) is overlaid in green.}
\label{fig:fig9_time-lag_maps}
\end{figure}

Let now focus on the cooling pattern in the light curves presented in Figure~\ref{fig:fig8_TNE_overview}.  For each pulse the peak in the different channels and 304 components appear sequentially. By choosing the timing of the rising of the peaks as a reference and not taking in account the multiple \SI{304}{\angstrom} peaks for the moment, it appears that  the intensity always peaks first in the hotter channels and then in the cooler channels, following the ordering of the peak temperature response of the channels: \SI{171}{\angstrom} ($\sim$ 0.8~MK), \SI{131}{\angstrom} ($\sim$ 0.5~MK) and cool \SI{304}{\angstrom} ($\sim$ 0.08~MK). The behavior of the extracted hot \SI{304}{\angstrom} ($\sim$ 1.8~MK) is less clear as it sometimes peaks in phase with the \SI{131}{\angstrom} channel. We have to take into account that contour (1) has two part: a western part that is  north-south oriented, and an eastern part oriented south-east. But here, we have averaged the light curve over the full contour, in which different pattern may mix. This was already hinted by the presence of rain only in the western part of contour (1) and will become clear later in this section (Figure~\ref{fig:fig9_time-lag_maps}). 
Nonetheless, we overall can conclude that this behavior is consistent with cooling observed in EUV channels. This is a property commonly observed in the X-ray and EUV corona \citep[e.g.,][]{ugarte-urraInvestigationVariabilityHeating2006, ugarte-urraActiveRegionTransition2009, winebargerCoolingActiveRegion2005, viallPatternsNanoflareStorm2011, warrenConstraintsHeatingHightemperature2011}. 

In order to investigate quantitatively and spatially the time-lags between the channels, we construct time-lag maps of the ROI using the cross-correlation method introduced in \citet{viallEvidenceWidespreadCooling2012}. For each pixel of the FOV, cross-correlations are performed between light curves of a pair of channels. The peak cross-correlation value gives the time lag between these two channels. For this analysis, we use the "rain dataset" (cadence of \SI{1}{\minute}), spatially binned to maximize the S/N.
We use the original \SI{304}{\angstrom} that is, as previously demonstrated, dominated by its cool component in the area covered by contour (1). We restrict our analysis  to the portion of the sequence between 26 hours and 50 hours after its start. This does not change our conclusions. However, it allows us to exclude the first two EUV pulses, which exhibit several \SI{304}{\angstrom} peaks for a single 12-hour period EUV pulse, as well as the fifth pulse during which coronal rain is no longer present and the structure has significantly rotated. We explore time shifts between the light curves from \SI{-300}{\minute} to \SI{300}{\minute}, which correspond to less than half of the periods and to typical time shifts explored in other studies \citep[e.g.,][]{fromentEvidenceEvaporationincompleteCondensation2015}.

In Figure~\ref{fig:fig9_time-lag_maps}, we show the time-lag maps for two pairs of channels: 171-131 and 171-304. We took the convention of a coronal plasma that would cool from the peak temperature response of the \SI{171}{\angstrom} channel,  which mean that positive time lags (in red) indicate cooling. We immediately notice that the two maps are dominated by unexploitable values. These are the black areas for which the peak cross-correlation value is below 0.2. This includes the eastern part of contour (1) for both pairs of channels.  
This can be due to the generally lower S/N off-limb, in particular here for the \SI{131}{\angstrom} and \SI{304}{\angstrom} channels. 

Nonetheless, some clear patterns emerge from the exploitable part of the maps. 
The 171-131 map is showing mostly zero time lags. We note, however, that given the time resolution chosen, we cannot resolve processes operating at a faster timescale than \SI{1}{\minute}. It is commonly observed in time-lags analyses and interpreted to be due to plasma not cooling below the peak response of the \SI{171}{\angstrom} channel or a very rapid cooling between the peak response of the \SI{171}{\angstrom} and \SI{131}{\angstrom} channels \citep[see Section~3.3 of][for an exhaustive discussion]{fromentMultiscaleObservationsThermal2020}. 
Zero 171-131 time lags are observed where no recurring coronal rain is observed which is consistent with the former explanation. Conversely, in the northern part of contour (1), where the rain falls back to the Sun surface, the 171-131 time lag is also close to zero. This could be due to plasma motions, that is the coronal rain falling, themselves.
There are, however, positive time lags in the "elbow" of contour (1) where coronal rain is appearing at first. In the western part of the contour, here taken as the union of contour(1) and a box delimited by $\rm{Solar-X} =[\SI{-970}{\arcsecond};\SI{-870}{\arcsecond}]$ and $\rm{Solar-Y} =[\SI{-430}{\arcsecond};\SI{-680}{\arcsecond}]$, we find a median time lag of \SI{10}{\minute}, with an Interquartile Range (IQR) equal to \SI{33}{\minute}, covering only the positive values. This mean the plasma is cooling from the \SI{171}{\angstrom} peak temperature response toward the \SI{131}{\angstrom} peak temperature response.
The 171-304 shows mostly positive time lags in contour (1). This mean the plasma is cooling from the \SI{171}{\angstrom} peak temperature response toward the 304 peak temperature response that is in the transition region. The median time lag in the western part of the contour(1) is \SI{69}{\minute} (IQR of \SI{70}{\minute} also only covering positive values). 
We observe that the reported values are quite sensitive to the cross-correlation threshold. We are thus not going to further comment the values themselves, only the ordering is relevant. Moreover, we observe a strong spatial coherence of this cooling signature.
To summarize: the time-lags values for contour (1) are consistent with previous studies of combined long-period EUV pulsations and coronal rain events, in particular the one of \citet{auchereCoronalMonsoonThermal2018}.
From these cooling signatures, combined with the periodicity of the signal and the recurring appearance of coronal rain, we can conclude that the present event has similar characteristic with events demonstrated to be TNE off-limb \citep[e.g.,][]{auchereCoronalMonsoonThermal2018, fromentMultiscaleObservationsThermal2020} and on-disk \citep[][except that rain showers were not investigated for these events.]{fromentEvidenceEvaporationincompleteCondensation2015}.

For contour (3,4,5), i.e. EUV pulsations linked to transient coronal events, no clear EUV cooling pattern was found. The dynamic of these rain showers and their location will be described in Sections~\ref{ss:rain_topo} and~\ref{ss:rain_dynamic}.

\section{Pseudo-streamer {topology and }dynamics and relationship with the TNE cycles}\label{s:PS_dynamic_and_TNE}

Section~\ref{s:PS_context} describes the long-term evolution of the EUV structures and establishes their association with a pseudo-streamer topology. In Section~\ref{s:TNE_signatures} we study the TNE cycles signatures in the EUV structures. In the present Section, we aim to connect the TNE cycles observations with the magnetic topology.

\subsection{Pseudo-streamer dynamics}\label{ss:PS_dynamic}

\begin{figure} 
\centerline{\includegraphics[width=\textwidth]{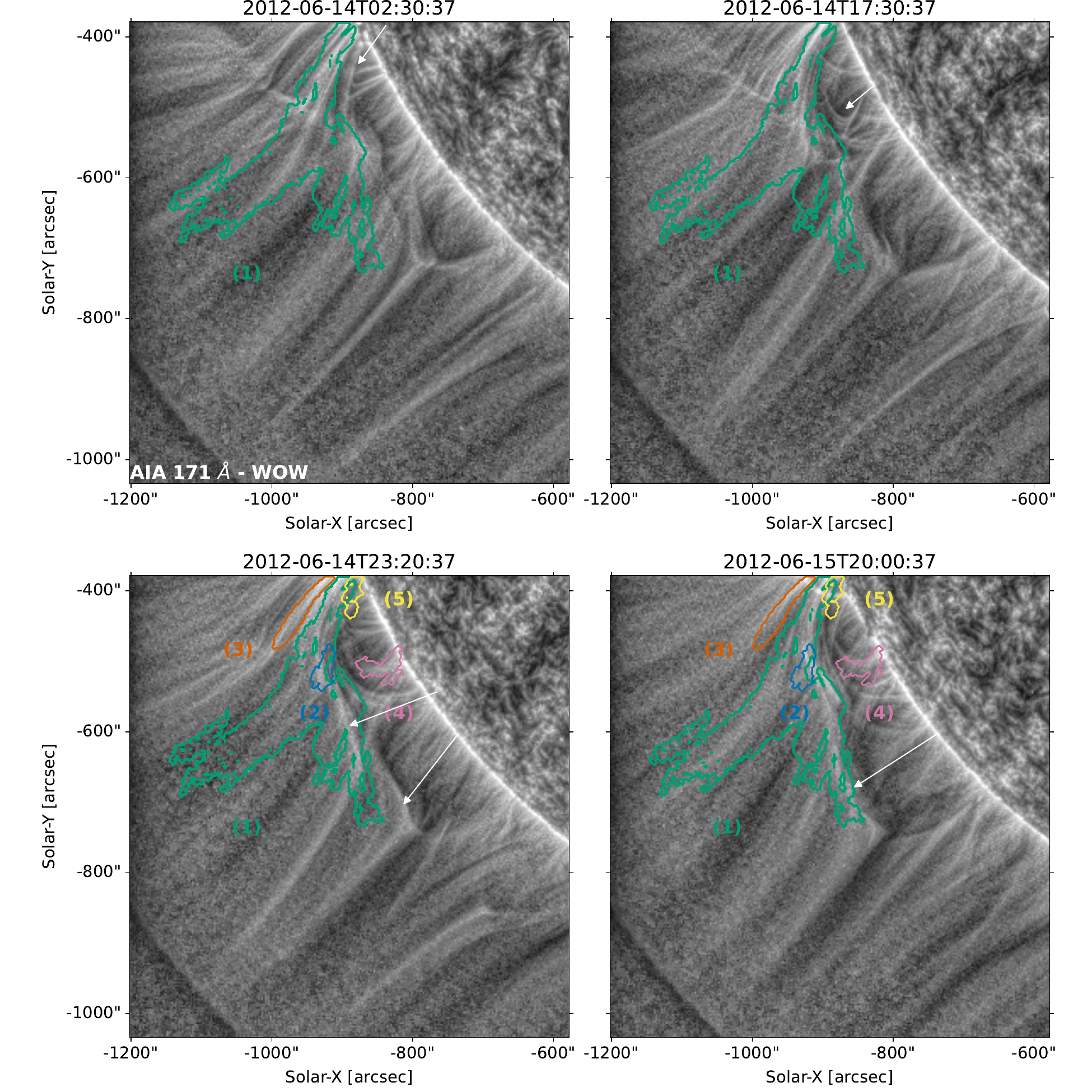}}
\caption{Four time steps in the \SI{171}{\angstrom} sequence (processed with WOW). Long-period EUV pulsations contours (1), (2), (3), (4) and (5) are overlaid in green, blue, orange, pink and yellow, respectively. The small contours are only included in the bottom panels for visibility. The white arrows indicate examples of: dynamical opening of the closed field (top left panel), dynamical closing of the open field (top right panel), and  bright EUV elbow-like structures that traces the intersection between the PS-dome and the open vertical fan (bottom panels). These behaviors are best observed in the animation accompanying the figure.}
\label{fig:fig10_TNE_open_structure}
\end{figure}

We know, from observations \citep{massonDynamicsTransitionCorona2014} and MHD modeling \citep{pellegrin-frachonInterchangeReconnectionDynamics2023}, that pseudo-streamer fine dynamics result from interchange reconnection between the closed field below the PS-dome and the surrounding open field. This leads to the dynamical opening and closing of the closed and open field, respectively. We point out to some examples of such dynamics, as seen from the EUV observations ("pulsation dataset" magnified with WOW), in Figure~\ref{fig:fig10_TNE_open_structure}. These examples are bright EUV lines or loops identified in the animation associated with Figure~\ref{fig:fig10_TNE_open_structure}. 
As an example, on June 14, at 02:20:37 UT, three bright loops are identified in the northern lobe, within the box defined by $\rm{Solar-X} =[\SI{-900}{\arcsecond};\SI{-850}{\arcsecond}]$ and $\rm{Solar-Y} =[\SI{-500}{\arcsecond};\SI{-400}{\arcsecond}]$ i.e. just below the northern part of contour (1) (see the white arrow in the top left panel of Figure~\ref{fig:fig10_TNE_open_structure}). Between 02:20:37 UT and 04:50:37 UT they rise slowly, reach the open-closed boundary and become open. Moreover, the opened bright structures above the northern and the middle lobes are not static but show some apparent slipping motions. Those two simultaneous behavior  indicates that field lines are opening up through interchange reconnection. 
At 17:10:37 UT, faint bright loops appear in the northern lobe. We clearly see its apex located near $\rm{Solar-X},\rm{Solar-Y} =[\SI{-870}{\arcsecond};\SI{-525}{\arcsecond}]$ (see the white arrow in the top right panel of Figure~\ref{fig:fig10_TNE_open_structure}). Between 17:10:37 UT and 18:00:37 UT, these new loops become brighter and shrink a bit, which is consistent with newly reconnected loops that closed down after interchange reconnection. Those two examples are not isolated cases. They illustrate the overall dynamics in this event, where bright loops are closing down or opening up all along the open-closed boundary as seen in EUV. 
This open-closed boundary can be seen as an EUV elbow-like structure (see the animation associated with Figure~\ref{fig:fig10_TNE_open_structure}). It resembles the shape of the open field lines that wraps above the PS-dome as seen in e.g. Figure~\ref{fig:fig4_PS_rotation_view}.
Here, it traces the intersection between the PS-dome and the open vertical fan, i.e. the closed separator where magnetic reconnection occurs \citep{parnellStructureMagneticSeparators2010, pellegrin-frachonInterchangeReconnectionDynamics2023}. With time, this bright EUV elbow-like structure seems to move southward which is most likely a direct consequence of the orientation change of the PS-dome caused by the solar rotation (see Section~\ref{fig:fig2_topology} and Figure~\ref{fig:fig4_PS_rotation_view}). Toward the middle of the sequence, we can witness the elbow-like structure simultaneously at two locations (see the two white arrows in the bottom lefts panel of Figure~\ref{fig:fig10_TNE_open_structure}). Toward the end of the sequence,
the only visible EUV elbow-like structure is further south, at $\rm{Solar-X},\rm{Solar-Y} =[\SI{-800}{\arcsecond};\SI{-750}{\arcsecond}]$ (see the white arrow in the bottom right panel of Figure~\ref{fig:fig10_TNE_open_structure}). 
The dynamics of the EUV structures in this pseudo-streamer topology clearly indicates that interchange reconnection occurs 
continuously and non-impulsively during the entire time of the observation and all along the open-closed boundary identified in the EUV observations.

\subsection{Location of the EUV pulsations and coronal rain showers in relationship with the magnetic topology}\label{ss:rain_topo}

By comparing the location of the EUV pulsations contours and the dynamics of the PS in this event (Figure~\ref{fig:fig10_TNE_open_structure}) we can conclude on the nature of the  magnetic field associated with those contours. Thus, contours (1), (2) and (3) include open-field or large loops. We notice that the western section of contour (1) follows almost perfectly the EUV open-closed boundary. The shape of this contour, constructed in the Fourier space as explained in Section~\ref{ss:pulsations_analysis}, is reflecting the changes of position of this boundary as the pseudo-streamer rotates. Contours (4) and (5) are detected in the loops below the PS-dome.

\begin{figure} 
\centerline{\includegraphics[width=\textwidth,clip=]{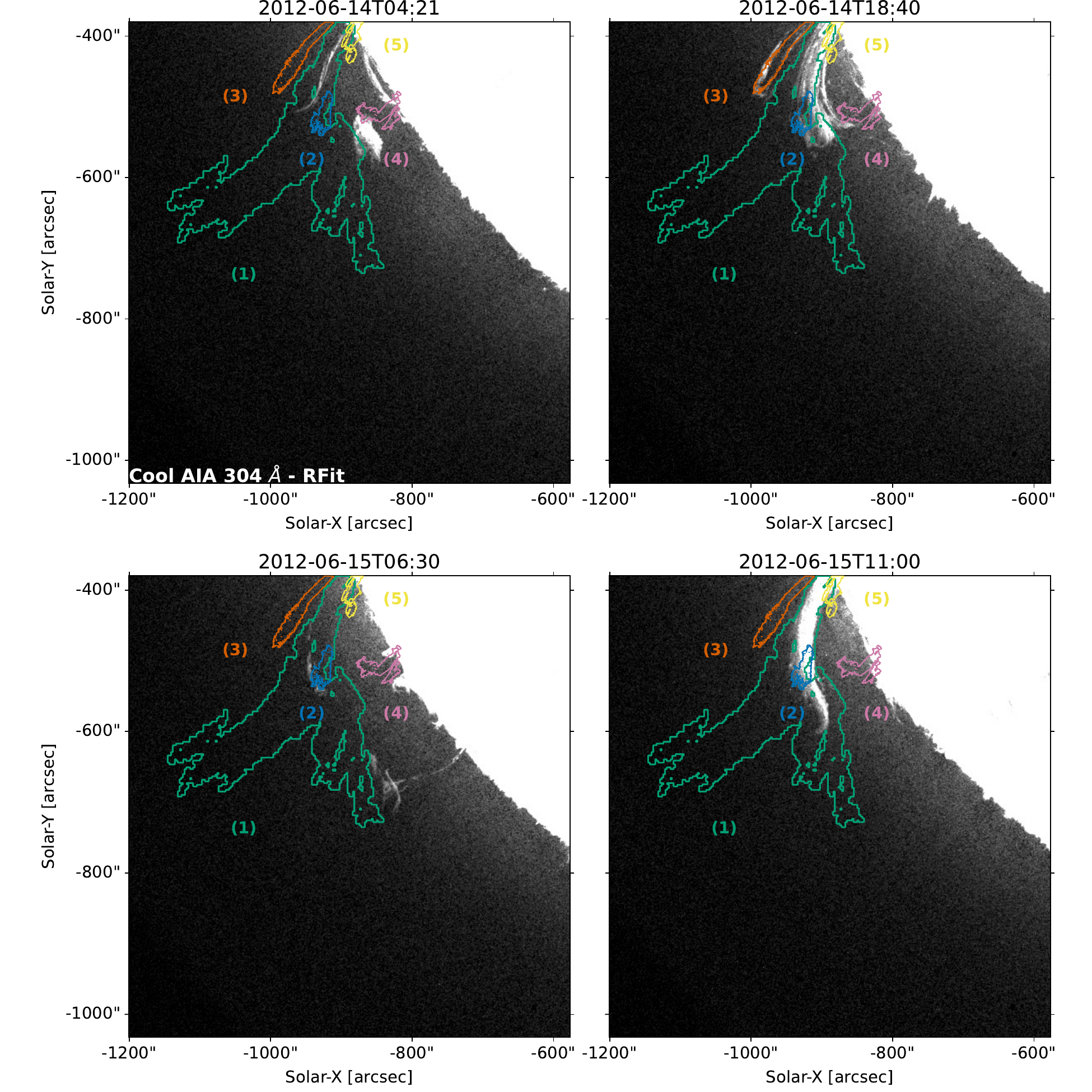}}
\caption{Different type of coronal rain paths observed during the event using cool \SI{304}{\angstrom} images extract using the RFit method. Long-period EUV pulsations contours (1), (2), (3), (4) and (5) are overlaid in green, blue, orange, pink and yellow, respectively. The four different snapshots show examples of the rain dynamics and morphology of the showers.  These events are best observed in the animation accompanying the figure.}
\label{fig:fig_11_rain_types}
\end{figure}

Let us now go back to the coronal rain showers dynamics, inside and outside of the EUV pulsations contours, and comment on the spatial location of the rain with respect to the field topology.
In Figure~\ref{fig:fig_11_rain_types}, we highlight examples of morphology and location of the coronal showers observed during the sequence. We refer the reader to the accompanying movie, for a full appreciation of the rain appearance and fall. We add the EUV pulsations contours whose position that can be used as markers of the PS topology/dynamics when comparing Figure~\ref{fig:fig_11_rain_types} and  Figure~\ref{fig:fig10_TNE_open_structure}, and the animations. Coronal rain is both seen in regions dominated by closed and open field, i.e. above and below the open-closed boundary as deciphered in the EUV images of Figure~\ref{fig:fig10_TNE_open_structure}. In the top left panel of Figure~\ref{fig:fig_11_rain_types} we see coronal rain, on June 14 at 04:21 UT, forming both below and above the open-closed boundary. A slender rain shower located in the northern part of contour (1) is curved at its tip, forming a hook-like feature. This does not look like a loop. We argue that this shower is forming in the open field. On the contrary, the rain clumps appearing in between contours (1) and (4) are likely related to the closed field. The rain forms in altitude and then falls to the surface toward both footpoint of the loops. In the top right panel of Figure~\ref{fig:fig_11_rain_types} (June 14 at 18:10 UT), we show a later example where most of the rain shower seem to be both related to open and close field regions, with rain forming inside and in the vicinity of contour (1) and then intersecting contours (4) and (5) during it fall to the surface. Inside contour (3), we see a fainter coronal rain shower. From its morphology it is not clear whether it is related to open or closed field or both. However, we can mention that toward the beginning of the sequence, mainly around  01:00 UT and  11:00 UT, we see similar showers in the close vicinity, but not inside, contour (3). These may be linked to the same bundle of field lines and differences may occur because of the solar rotation. In the bottom left panel of Figure~\ref{fig:fig_11_rain_types}, we show an example of rain showers (June 15 at 06:30 UT) that also appear in the southern part of the pseudo-streamer, i.e. in the middle EUV lobe. These rain clumps appear outside of the contour and thus do not affect the time series seen in e.g. Figure~\ref{fig:fig8_TNE_overview}. This may be related to the PS-dome presenting differently because of the solar rotation as it was noticed in Section~\ref{s:PS_dynamic_and_TNE}. Conversely, some faint rain clumps also appear in the same area on June 14 between 07:30 and 10:30 UT. 
Finally, in the bottom right panel of Figure~\ref{fig:fig_11_rain_types}, we show an example of one of the main recurrent rain shower path, here on June 15 at 11:00 UT. This strong coronal rain shower appears within contour (1) and, as we detailed in Section~\ref{ss:rain_analysis} and Figure~\ref{fig:fig8_TNE_overview}, is linked to the main \SI{304}{\angstrom} peaks in this contour. At this particular time we clearly see the hook-like feature at the tip of the rain path. The rain is falling toward a single footpoint. Again, we argue this is a clear occurrence of coronal rain formation in the open field. It resembles the shape of the elbow-like structure as seen in e.g. Figure~\ref{fig:fig10_TNE_open_structure}. Moreover, we notice that this path is similar to the cooling area (in red) described in Figure~\ref{ss:cooling_analysis}. This area likely marks the coronal rain formation region.
However, for earlier and later occurrences, e.g. between June 14 at 23:00 UT and June 15 at 04:00 UT and on June 15 around 20:00 UT, the hook-like feature does not appear or at least not that clearly. Especially for these later time it is more difficult to conclude if the rain is exclusively formed in the open field regions or in both open and closed fields.

\subsection{Imprint of the coronal rain showers in the light curves}\label{ss:rain_dynamic}

We will now briefly go back to the link between the appearance of coronal rain showers with the EUV light curves variability, in particular for \SI{304}{\angstrom}. In   Sections~\ref{ss:rain_analysis} and~\ref{ss:cooling_analysis} we already covered the topic for contours (1) and (2). We may further note that the shorter period detected in the light curves of contour (2) (Figure~\ref{fig:fig7_additionnal_contours}) seems to be close to the coronal rain showers finer timescale noticed already in the contour (1) light curves, that is multiple \SI{304}{\angstrom} peaks for a single \SI{171}{\angstrom} pulse (Figure~\ref{fig:fig8_TNE_overview}). This shorter timescale may be linked to the coronal rain formation process itself or to superimposition of several structures on the LOS.

In the light curves of contour (3) (Figure~\ref{fig:fig7_additionnal_contours}), we notice a large \SI{304}{\angstrom} peak from approximately 18 to 20 hours after the start of the sequence. This is linked to the coronal rain shower we discussed in Section~\ref{ss:rain_topo} with Figure~\ref{fig:fig_11_rain_types}. This \SI{304}{\angstrom} peak appears after the \SI{171}{\angstrom} peak (Figure~\ref{fig:fig7_additionnal_contours}). However, from a single transient rain event it is impossible to conclude on a possible recurring cooling pattern and the evidence of TNE. Moreover, as already noticed, there are likely projections effects which may be amplified by the slenderness of contour (3).
In the light curve of contour (4) (Figure~\ref{fig:fig7_additionnal_contours}), we see many \SI{304}{\angstrom} peaks. However, contrary to what we observed for contour (1) and (2), these do not seem to be systematically positively delayed compared to the \SI{171}{\angstrom} peaks. The \SI{304}{\angstrom} peaks appear either: at the same times as the \SI{171}{\angstrom} peaks (e.g. at about 26 hours after the start of the sequence), before the \SI{171}{\angstrom} peaks (e.g. in between 4 and 10 hours after the start of the sequence), or uncorrelated with the \SI{171}{\angstrom} peaks (e.g. from 40 hours after the start of the sequence). Here again, we do not find a clear cooling pattern and thus TNE signatures. This may be due to LOS superimposition or the combination of the TNE and interchange reconnection processes. Indeed, contour (4) is clearly located in the area where loops are closing due to the interchange reconnection (see the accompanying animation of Figure~\ref{fig:fig10_TNE_open_structure}). Interestingly, we note, that the \SI{304}{\angstrom} peaks (and linked coronal rain showers), occurring in between 4 and 10 hours after the start of the sequence, are quasi-synchronous with the \SI{304}{\angstrom} peaks (and linked coronal rain showers) in the contours (1) and (2) light curves. This may indicate that the coronal rain showers occurring in the small loops crossing the area of contour (4) may come from the open field region and flow along the newly reconnected loops as proposed by \citet{liCoronalCondensationsCaused2018}. One other example may be seen from 17 hours after the start of the sequence as a coronal rain shower in contour (4) (see the accompanying animation of Figure~\ref{fig:fig_11_rain_types}) starts quasi-synchronously with the dynamical closing of field lines we pointed out in Figure~\ref{fig:fig10_TNE_open_structure} and in Section~\ref{ss:PS_dynamic}.
Deciphering in details this dynamics would require more investigation, which is beyond the scope of this paper. In particular, the interplay between the TNE cycles and the interchange reconnection process should be investigated further.
Finally, there is a high variability if the \SI{304}{\angstrom} light curve of contour (5) (Figure~\ref{fig:fig7_additionnal_contours}). This contour is likely located at the footpoint of both open and closed structures. In the accompanying animation of Figure~\ref{fig:fig_11_rain_types} we see that coronal showers coming from multiple structures fall in this area. Also, this contour is very close to the surface, such as a part of contour (4). The \SI{304}{\angstrom} light curves of both these contours are thus also affected by dynamics of low lying structures such as spicules or small prominences.

\section{Summary and Discussion}\label{s:Conclusion}

In this paper, we study the long-term EUV dynamics linked to a pseudo-streamer as seen off-limb in SDO/AIA images.
For 2.5 days, we tracked the pseudo-streamer structure that can be mainly observed in the \SI{171}{\angstrom}, \SI{131}{\angstrom}, and \SI{304}{\angstrom} SDO/AIA channels. We used PFSS reconstruction to model the global magnetic field topology of the region and reconciled it with the EUV \SI{171}{\angstrom} evolution. We can summarize our findings as follows.
The evolution of the EUV structures is consistent with the dynamics of a pseudo-streamer, in which interchange reconnection occurs throughout the observation, involving the dynamical opening and closing of the closed and open field, respectively.
In addition, the pseudo-streamer has the particularity that its orientation changes with the Carrington longitude which complicates the EUV LOS evolution. 
We detected long-period EUV pulsations with periodicities of a few hours. In the main pulsation contour, where 12-hours pulsations are detected, we 
further detected recurrent coronal rain showers in the \SI{304}{\angstrom} channel, that coincide in space and time with the pulsations in the hotter channels. 
The periodic EUV pulses appear sequentially in the different channels, following
the ordering of the peaks in their temperature response, with the coronal rain appearing by the end of the cycles. These cooling signatures combined with the periodic behavior in the AIA channels and the quasi-synchronous rain showers are strong evidence for TNE cycles. By further examining the dynamics of the pseudo-streamer in EUV, we identified that these cycles occur, not only in the closed field domain below the PS-dome but also in the region dominated by the open field and the large closed loops, both surrounding the PS-dome. Some rain showers are spotted in closed field which may come from usual TNE processes, as some are linked to additional EUV pulsations contours, or the release of condensations through magnetic reconnection at the pseudo-streamer closed separator. The later is hinted by the combined coronal showers dynamics and dynamical closing of open structure as deciphered in the EUV observations.

We note that the main period detected (12 hours) is in the upper range of periodicity reported in the statistical studies of \citet{auchereLongPeriodIntensityPulsations2014} and \citet{fromentPulsationsDintensiteLongue2016}. 
Previous studies \citep[e.g.][]{fromentOccurrenceThermalNonequilibrium2018} showed that longer TNE periods are generally linked to longer structures. This is compatible with the fact that we found the EUV pulsations to occur in the open field or in very large loops. Conversely, the additional pulsation events we detected have periods ranging from 2.9 to 4.6~hours and seem to occur in the open field and closed field area. Other parameters such as the volumetric heating strength \citep[e.g.][]{fromentOccurrenceThermalNonequilibrium2018} and the heating timescale \citep[e.g.][]{johnstonEffectsNumericalResolution2019} can affect the TNE period, if all these events are indeed linked to TNE. See \citet{antolinMultiScaleVariabilityCoronal2022}, for a detailed discussion.
It is also worth noting that about $50~\%$ of the long-period EUV pulsations events reported on-disk are found in the quiet Sun \citep[][]{auchereLongPeriodIntensityPulsations2014, fromentPulsationsDintensiteLongue2016}. These tend to have longer periods compared to the active regions events. These on-disk events have not yet been studied further, but we speculate that some of them may be related to the kind of events reported in the present paper.

We may now discuss our results and interpretations, and their possible implications. 
There are many similarities between our events and the ones previously reported by \citet{masonObservationsSolarCoronal2019} and \citet{liCoronalCondensationsCaused2018, liRepeatedCoronalCondensations2019, liRelationCoronalRain2020}. These events are located at open-closed topologies and more specifically in pseudo-streamers. Moreover, all these events were reported as recurrent, even though the periodicity of the events was not explored in these studies. 
\citet{liCoronalCondensationsCaused2018, liRepeatedCoronalCondensations2019, liRelationCoronalRain2020} events showed also similar EUV cooling signatures in the \SI{171}{\angstrom}, \SI{131}{\angstrom}, and \SI{304}{\angstrom} channels of AIA. Into more details we noticed that the positive 171-131 time lags (see Figure~\ref{fig:fig9_time-lag_maps}), are seen in the elbow-like structure, that traces the intersection between the PS-dome and the open vertical fan, and that encompasses the long-period EUV pulsations contour (1). This could indicate that this area is the forming region of the rain. This is again quite similar to the events reported by \citet{liCoronalCondensationsCaused2018, liRepeatedCoronalCondensations2019, liRelationCoronalRain2020}. 

\citet{masonObservationsSolarCoronal2019} concluded that TNE or interchange reconnection could explain their observation and that both processes could be present. With our study, we confirm that TNE cycles can happen in pseudo-streamers, with interchange reconnection dynamical opening and closing of field lines happening in parallel. 
Moreover, some of the long-period pulsations are detected in the open field which may be the open field that has newly reconnected. This was one of the hypotheses of \citet{masonObservationsSolarCoronal2019} that coronal rain forms in the newly open flux, filled with the hot and dense plasma formerly present in the closed flux. 
However,  we would argue that the periodicity of the cooling cycles can not be explained alone by the interchange process. The hour-long timescale is compatible with supergranular turnover timescale which may drive the pseudo-streamers dynamics \citep[e.g.][]{aslanyanEffectsPseudostreamerBoundary2021}, but we clearly see in the present event that the interchange reconnection occurs continuously. Conversely, these reconnection events may cause continuous heating across the PS-dome, i.e., at high altitudes, which needs to be understood in the context of TNE (i.e., footpoint heating).
Further observational studies and modeling are needed to understand the interplay of TNE and interchange reconnection.

Recent numerical works have also brought new perspectives.
Using 3D radiative MHD simulations, \citet{kohutovaSelfconsistent3DRadiative2020} studied the self-consistent formation of coronal rain in coronal loops. They also reported the formation of coronal rain in open flux present in the simulation. This is further supported by the dedicated 1D multi-fluid hydrodynamic simulations of \citet{scottSimulationThermalNonequilibrium2024}, that shows that TNE can occur in open field lines and thus directly affect the solar wind dynamics. 
There is an increasing interest for the numerical study of TNE in open-field regions or at open-closed boundaries. \citet{schlenkerEffectThermalNonequilibrium2021} used a 2.5D MHD code to study TNE in helmet streamers, i.e. the last closed loops connecting the two coronal holes of opposing polarity. They showed that TNE, in addition to the plasma dynamics, can significantly drive the magnetic dynamics of the streamers, especially near their top where the plasma $\beta$ is close to one. 
They find that the interplay between TNE and the magnetic field plays a dominant role in the observed dynamics. Even though coronal rain has never been reported yet in helmet streamers, these results bring interesting perspectives. Using also 2.5D MHD simulations, \citet{johnstonFilamentMassLosses2025} investigated the role of magnetic reconnection at a coronal null-point topology in which a cloud filament forms \citep[close to the observational case reported by][]{chenCoronalCondensationSource2022a}. They found that the formation of such cloud prominence, where condensation accumulates above a null-point, can be explained by evaporation-condensations cycles, i.e. TNE processes. They also concluded that while some condensations can be released through the null-point, magnetic reconnection does not play a direct role in the formation of the condensations. Such processes may explain some of the coronal rain showers in the PS-dome as described in Section~\ref{ss:rain_topo}.

All these numerical works show that TNE can also happen in open field and at open-closed boundaries. TNE manifestations provide strong constrains on the heating mechanisms in the corona as the evaporation-condensations cycles are driven by a quasi-steady and stratified heating.
\citet{sahinPrevalenceThermalNonequilibrium2022} studied coronal rain showers within an active region as seen off-limb and estimated that TNE volume is on the order of the active region volume itself. 
With our study, we reaffirm that TNE is also of interest at the sources of the solar wind, the spine of the northern pseudo-streamer we studied extend toward a few solar radii. The interplay of TNE and interchange reconnection could play a role in driving some of the solar wind dynamics.

%
\begin{acks}
We thank the anonymous referee for their careful reading of our work and their constructive comments.
We would like to give special thanks to Helle Bakke for her dedicated assistance in mining our long-period EUV pulsations catalog and for identifying events outside of the previously in place categories, and Jérôme Roudil for his further assistance in the events identifications.
We also thank Nina Bizien for her initial assistance on the PFSS modeling and plotting, and Daniel Nóbrega-Siverio for his help for making the time-distance figure. We thank the participants of the Coronal Cooling conference and Tom Van Doorrseleare for their insightful comments.
CF benefited from the excellent discussions among the members of the ISSI Bern International Team project $\#545$ (“Observe Local Think Global:
What Solar Observations Can Teach Us about Multiphase
Plasmas across Physical Scales”).

\end{acks}

\begin{authorcontribution}
This work was conceptually initiated by CF who also performed all the data analysis and produced the figures. SM enabled the identification of the pseudo-streamer, helped in the interpretation of the results, and is also the main contributor of the writing of Sections 2.2, 2.3 and 4.1.
CF and SM edited together the final draft.
\end{authorcontribution}

\begin{fundinginformation}
This research was supported by the Agence Nationale de la Recherche (ANR) for the CROSSWIND project under the grant ANR-24-CE31-2993 and the Centre National d’Études Spatiales (CNES), through its APR program. 
This research also benefited from the support of the International Space Science Institute (ISSI) in Bern. This work was also supported by the Action Thématique Soleil-Terre (ATST) of CNRS/INSU PN Astro, co-funded by CNES and CEA.
\end{fundinginformation}

\begin{dataavailability}
The SDO/AIA are available courtesy of NASA/SDO and the AIA science teams. SDO is part of NASA’s Living With a Star Program. This work used data provided by the MEDOC data and operations center (CNES/CNRS/Univ. Paris-Saclay), \href{http://medoc.ias.u-psud.fr/}{http://medoc.ias.u-psud.fr/} (DOI: \href{https://doi.org/10.48326/idoc.medoc.sdo.aia}{https://doi.org/10.48326/idoc.medoc.sdo.aia}). This work utilizes data produced collaboratively between Air Force Research Laboratory (AFRL) \& the National Solar Observatory (NSO). The ADAPT model development is supported by AFRL. The input data utilized by ADAPT is obtained by NSO/NISP (NSO Integrated Synoptic Program).
\end{dataavailability}

\begin{codeavailability}
This analysis relies on the \texttt{pfsspy} implementation of PFSS \citet{stansbyPfsspyPythonPackage2020} in Python.
This research also used version 0.7.2 of \texttt{aiapy} and  version 5.1.5 (DOI: \href{https://doi.org/10.5281/zenodo.591887}{https://doi.org/10.5281/zenodo.591887}) of the \texttt{SunPy} open source software package \citep{barnesSunPyProjectOpen2020}.
This research made use of wcsaxes, an open-source plotting library for Python hosted at https://wcsaxes.readthedocs.io/en/latest/. This research made also use of Astropy, a community-developed core Python package for Astronomy (astropy) (v3.1.0; The Astropy Collaboration et al. 2018); matplotlib, a Python library for publication quality graphics (matplotlib) (v3.0.2; Hunter 2007; Caswell et al. 2018); NumPy (numpy) (v1.15.4; Harris et al. 2020); seaborn (v0.9.0; Waskom et al. 2018); scipy (v1.1.0; Virtanen et al. 2020), and SolarSoftware (Freeland and Handy 1998). We also used the scientific color maps of Crameri, F. (2018) DOI: \href{https://doi.org/10.5281/zenodo.1243862}{https://doi.org/10.5281/zenodo.1243862}).
\end{codeavailability}

\begin{ethics}
\begin{conflict}
The authors declare that they have no conflicts of interest.
\end{conflict}
\end{ethics}

%
%
\bibliographystyle{spr-mp-sola}
\bibliography{references.bib}

\end{document}